\documentclass[manuscript]{acmart}
\usepackage[utf8]{inputenc}
\usepackage{subfigure}

\def\plaintitle{Understanding Visual Saliency in Mobile User Interfaces}
\def\plainauthor{Luis A. Leiva, Yunfei Xue, Avya Bansal, Hamed R. Tavakoli, Tuğçe Köroğlu, Niraj R. Dayama, Antti Oulasvirta}
\def\plainkeywords{Human Perception and Cognition; Interaction Design; Computer Vision; Deep Learning}

\newcommand{\tth}[1]{\textbf{\small#1}}

\hypersetup{
  pdftitle={\plaintitle},
  pdfauthor={\plainauthor},
  pdfkeywords={\plainkeywords},
}

\makeatletter
\if@twocolumn
  
\else
  
\fi
\makeatother

\copyrightyear{2020}
\acmYear{2020}
\setcopyright{acmlicensed}
\acmConference[MobileHCI '20]{22nd International Conference on Human-Computer Interaction with Mobile Devices and Services}{October 5--8, 2020}{Oldenburg, Germany}
\acmBooktitle{22nd International Conference on Human-Computer Interaction with Mobile Devices and Services (MobileHCI '20), October 5--8, 2020, Oldenburg, Germany}
\acmPrice{15.00}
\acmDOI{10.1145/3379503.3403557}
\acmISBN{978-1-4503-7516-0/20/10}

\begin{document}

\title{\plaintitle}

\author{Luis A. Leiva}
\affiliation{
  \institution{Aalto University}
  \country{Finland}}
\email{firstname.lastname@aalto.fi}

\author{Yunfei Xue}
\affiliation{
  \institution{Aalto University}
  \country{Finland}}
\email{firstname.lastname@aalto.fi}

\author{Avya Bansal}
\affiliation{
  \institution{Indian Institute Of Technology, Goa}
  \country{India}}
\email{avya.bansal.16001@iitgoa.ac.in}

\author{Hamed R. Tavakoli}
\affiliation{
  \institution{Nokia Technologies}
  \country{Finland}}
\email{hamed.rezazadegan_tavakoli@nokia.com}

\author{Tuğçe Köroğlu}
\affiliation{
  \institution{Yildiz Technical University}
  \country{Turkey}}
\email{krgl.tugce@gmail.com}

\author{Jingzhou Du}
\affiliation{
  \institution{Huawei Technologies}
  \country{China}}
\email{dujingzhou@huawei.com}

\author{Niraj R. Dayama}
\affiliation{
  \institution{Aalto University}
  \country{Finland}}
\email{firstname.lastname@aalto.fi}

\author{Antti Oulasvirta}
\affiliation{
  \institution{Aalto University}
  \country{Finland}}
\email{firstname.lastname@aalto.fi}

\renewcommand{\shortauthors}{L. A. Leiva et al.}

\begin{abstract}
For graphical user interface (UI) design, it is important to understand what attracts visual attention. While previous work on saliency has focused on desktop and web-based UIs, mobile app UIs differ from these in several respects. We present findings from a controlled study with 30 participants and 193 mobile UIs. The results speak to a role of expectations in guiding where users look at. Strong bias toward the top-left corner of the display, text, and images was evident, while bottom-up features such as color or size affected saliency less. Classic, parameter-free saliency models showed a weak fit with the data, and data-driven models improved significantly when trained specifically on this dataset (e.g., NSS rose from 0.66 to 0.84). We also release the first annotated dataset for investigating visual saliency in mobile UIs.

\end{abstract}



\maketitle

\section{Introduction}
\label{sec:introduction}
For a graphical object, the notion of visual saliency refers to the ability to attract visual attention,
for the given visual properties of that object and the rest of the display~\cite{Borji13}.
In practice, attention is drawn to \textcolor{red}{\textbf{visually unique stimuli}}.
Through use of color and boldface, the three words above
stand out relative to the rest of this paragraph.
Regions and objects that are unique in terms of visual primitives
-- such as size, color, shape, orientation, or motion -- tend to stand out~\cite{Li15, Wolfe04}.
The biological basis for this phenomenon is well-known~\cite{Tsotsos91}:
saliency emerges in parallel processing of retinal input at lower levels in the visual cortex~\cite{Veale17}.
From mature-level research on saliency has sprung a wealth of applications in visual computing,
scene classification~\cite{Siagian07},
video summarization~\cite{Ma05}, image segmentation~\cite{Mishra09} and compression~\cite{Ouerhani01},
object detection~\cite{Ehinger09}, and other areas.
Nevertheless, predicting where people look at is paradigmatically more ambiguous
than typical computer vision tasks such as image segmentation or object detection,
since users weight visual features differently when deciding where to look~\cite{Zhao13}.

User interface (UI) designers examine visual saliency to understand what users will be drawn to when seeing a display,
and to avoid designs that appear cluttered~\cite{Rosenholtz11, Still10}.
Inexpensive commodity eye-trackers have made empirical data collection more popular.
However, while some findings have been reported for desktop- and web-specific UIs,
no published empirical research on visual saliency has focused on mobile ones.
This is alarming, sicne these devices are among today's most prevalent computing terminals.
To our knowledge, the only study in this area is one by Gupta et al.~\cite{Gupta18},
who used crowdsourced data to validate computational models on mobile UIs,
but did not report factors affecting saliency.

\begin{figure}[!ht]
\centering
\includegraphics[width=0.8\linewidth]{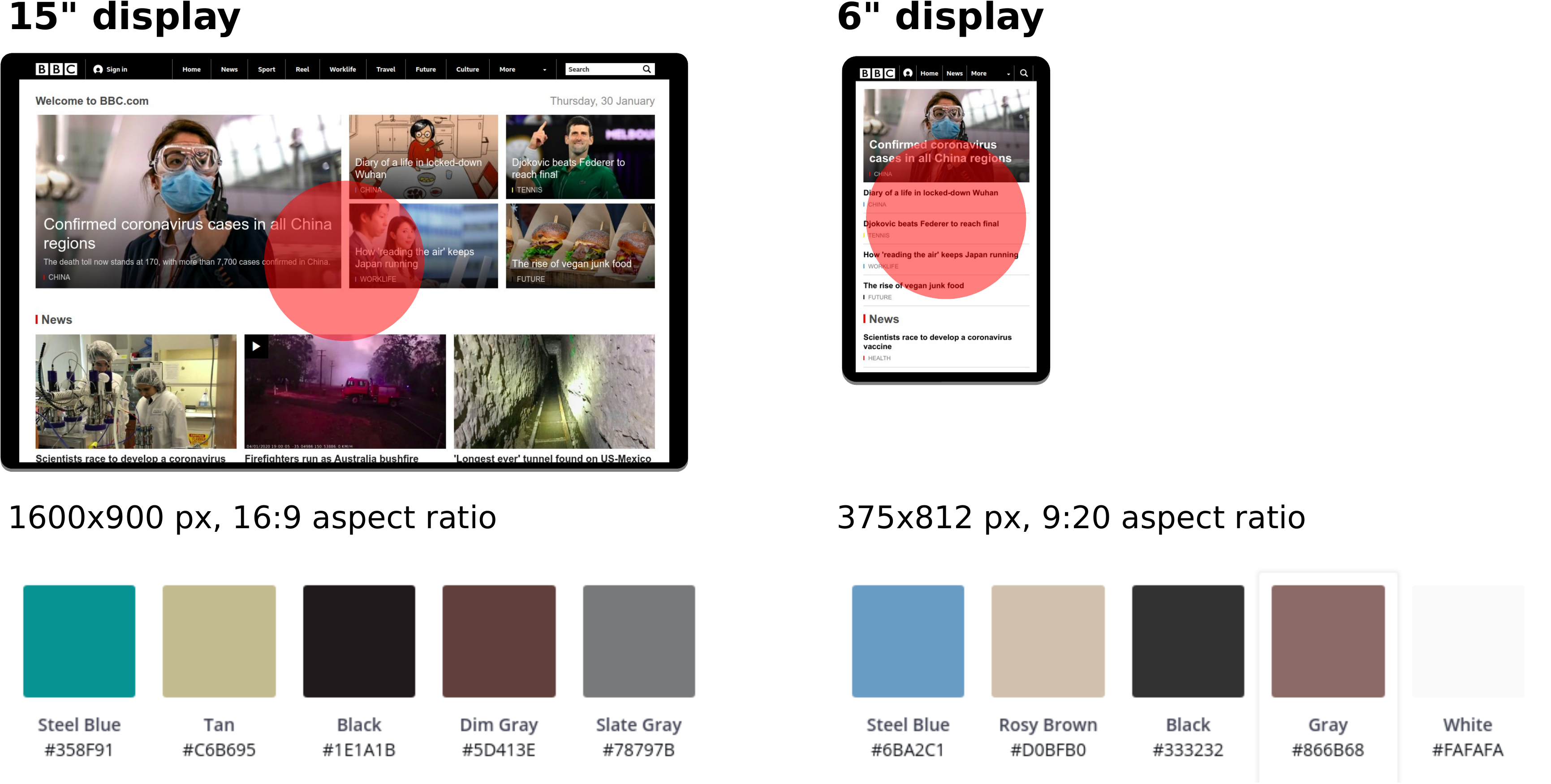}
\caption{A UI for a mobile device can differ in visual features' distribution from the equivalent UI on a laptop computer.
Shown here are the screen resolution, aspect ratio, and color distributions (according to \url{https://www.canva.com/colors/}) of a sample webpage.
Red circles represent areas of foveal vision in viewing the UIs from a 60\,cm distance.}
\Description{A desktop and a mobile UI are shown side by side, to better notice their perceptual differences.}
\label{fig:comparison}
\end{figure}

Xu et al.~\cite{Xu12touch} (with follow-up by Ni et al.~\cite{Ni14touch})
investigated \emph{touch saliency} for mobile devices
as a method estimating saliency from touch points.
However, touch points are not tantamount with visual saliency.
Moreover, the experimental stimuli used were outdoor scenes, not UIs.
Later on, Xu et al.~\cite{Xu16_chi} predicted saliency for desktop UIs,
taking as input the users' mouse and keyboard actions,
but this setup cannot be transferred to mobile UIs.
While Shen et al.~\cite{Shen14} used an \emph{ad-hoc} model for predicting visual saliency of webpages,
webpages as viewed on a laptop computer differ from mobile UIs in several respects; see \autoref{fig:comparison}.

While there have been many findings on visual saliency in non-mobile UIs
-- e.g., related to color bias, face bias, horizontal bias, etc. (see `\nameref{sec:related-work}' section) --
they cannot be expected to hold automatically for mobile UIs,
for several reasons:
\text{(1)}~Mobile displays are much smaller than desktop or laptop displays,
which means that, everything else being equal, fewer objects compete for attention.
\text{(2)}~Mobile devices span a much smaller field of view.
Thus, graphical objects tend to lie closer to the fovea.
\text{(3)}~Mobile displays are often vertically oriented,
which entails less use of the horizontal dimension of peripheral vision.
\text{(4)}~Designers create UI designs specifically for mobile devices,
and mobile UIs may feature a different distribution of visual primitives.
For example, interactive elements such as buttons are relatively large in comparison to their desktop versions.
\text{(5)}~Users may have learned different strategies for looking at mobile UIs,
reflecting the visual characteristics or underlying tasks involved.
As the viewport and the distribution of visual primitives change, so does saliency.
\autoref{fig:comparison} shows an example comparison of screen size and
color distribution between a typical laptop and mobile device.
In sum, what may be salient in a desktop UI may be less or more so in its mobile analogue.

Computational models too need to be reassessed for this special context.
In recent decades, plausible and accurate computational models of visual saliency have emerged.
Early (``classic'') models~\cite{Itti98, Parkhurst02} were based on biologically motivated visual primitives (color, shape, etc.),
followed by center--surround operations highlighting local changes,
which were combined into a ``conspicuity map''~\cite{Harel07, Itti98}.
In the last five years, data-driven models, most recently using deep learning,
have surpassed these in predictive accuracy~\cite{He16}.
However, deep learning models are trained on particular visual features
and may fail when the underlying distribution changes.
Critical validation of computational models is important,
since they could play a strong role in interface design~\cite{Rosenholtz11}.
We see two main uses of these models:
An \emph{inverse approach} starts with behavioral data, such as scrolling or cursor movement logs,
and produces heatmaps of the UI regions that are most likely attended~\cite{Xu16_chi}.
A \emph{forward approach} takes a display as input and produces a fixation map depicting predictions as to the most salient areas~\cite{Cornia18}.
Using a sound computational model of visual saliency,
designers can make changes to a UI layout and predict the effects without having to run user studies.

This paper presents the first empirical investigation of statistical effects of visual saliency in mobile UIs
and validation of several well-known computational models,
including state-of-the-art deep learning models,
from high-quality eye-tracking data.
We collected the data for a large number of representative (stock) mobile interfaces
viewed from a distance typical for a mobile device,
using a high-fidelity stationary eye-tracker that is calibrated regularly to minimize drift.
We studied the Chinese mobile ecosystem because it is currently the largest one,\footnote{
  See e.g. https://www.statista.com/topics/1416/smartphone-market-in-china/}
though our results should be easy to replicate in other cultures,
given that bottom-up saliency is a general phenomenon of attention.
In summary, the contribution of this paper is threefold:
(1)~we assess known saliency-related phenomena with mobile devices,
among them biases toward learned locations and features;
(2)~we report several validation metrics for models of visual saliency,
showing that state-of-the-art deep learning models can improve significantly in accuracy
when trained on a mobile-specific dataset; and
(3)~we release the first dataset for visual saliency of mobile UIs,
which includes fixation data on 193 UIs annotated with per-element labels and bouding boxes.

\section{Related Work}
\label{sec:related-work}
Visual saliency is the perceptual quality of some objects or regions attracting attention by standing out from the rest of the view~\cite{Itti98}.
Saliency has a role in the control of visual attention,
especially in picking the next fixation points~\cite{Itti:2007, Richard}.
Below, we review the empirically known effects of saliency,
which we assessed in the case of the new dataset,
and the state of computational modeling in this area.

\subsection{Empirical evidence on saliency}

What is salient is affected by both \emph{bottom-up} and \emph{top-down} factors.
The former are characterized by uniqueness in visual primitives of the stimulus,
such as color, shape, size, orientation, or motion.
An object that is unique in these respects tends to draw attention.
For instance, in an image full of green tones and green-filled shapes,
if a color such as red appears, observers tend to look at the red shape~\cite{Lim2012}.
Top-down factors include task goals and expectations
that are based on the learned statistical distribution of features; see e.g., \cite{series2013learning}.
For example, in many natural scenes that show a horizon, most information lies close to the horizontal medial line,
which also attracts attention.
When the visual task or content changes, both bottom-up and top-down factors may change.
Therefore, a saliency-related empirical effect reported for one context does not trivially carry over to another.

Prior research looking at saliency of natural scenes has found several replicated effects, or biases,
which we revisit in this paper:

\emph{1. Center bias:}
Studies have reported a bias toward looking at the center of the screen when viewing natural scenes~\cite{Henderson, Nuthmann}.
This is supposedly driven by the statistical distribution of image features in such scenes~\cite{Tatler2007, Tseng09}.
However, the effect has been replicated with other media, such as video~\cite{Marat}, text~\cite{Rayner}, and single objects~\cite{Nuthmann}.
Whether this holds for mobile UIs is unclear,
since much of their most informative elements lie in the upper half of the UI.
That said, gazing in the middle might yield the best overview of the UI for peripheral vision.

\emph{2. Horizontal bias:}
In looking at natural images featuring objects,
fixations tend to be distributed more horizontally than vertically~\cite{Nuthmann, Ossandon}.
Again, mobile UIs differ from natural scenes
in that they organize the information vertically, not horizontally.
Therefore, we might see this effect weaken.

\emph{3. Color bias:}
Thus far, color brightness and contrast have been counted among the primary features driving bottom-up saliency~\cite{Etchebehere17, Hamel14}.
Mobile UIs typically contain colorful icons and images
that may be perceived as highly salient;
therefore, we would expect this effect to hold.

\emph{4. Text bias:}
Perhaps due to the importance of textual materials in lived environments,
a bias toward textual elements has been reported~\cite{Humphrey2012, Wang2012}.
When an observer is told not to look at text, initial fixations take longer~\cite{Cerf2009}.
In mobile UIs, text plays a similarly important role, also in icons, labels, headings, logos, etc.
This effect may, therefore, still exist.

\emph{5. Face bias:}
Perhaps because of evolutionary advantages,
attention is drawn to human faces~\cite{Cerf2009, Ehinger09}.
Initial fixations have been found to take longer when the user is asked to avoid faces~\cite{Cerf2009}.
Mobile UIs often include graphics that may contain faces.
However, faces play a less informative role in mobile UIs,
because they are typically not interactive elements.
Hence, this effect might be expected to decrease or disappear.

\subsection{Computational saliency models}

Given a stimulus image, a computational model of visual saliency
produces a density map showing the amount of conspicuity --
that is, how much a pixel stands out against other pixels.
In this paper, we compare two model types.
(1)~\emph{Bottom-up models}, or stimulus-driven,
are based on visual primitives such as color, size, and shape~\cite{Borji13, Borji_2013_ICCV}.
These models can be expected to work well in the case of a user shown a UI for the first time for free viewing~\cite{Frintrop10, Itti98}.
(2)~\emph{Data-driven models} are models that, though make predictions based on image features~\cite{Borji19},
are trained using eye-movement data and may have architectural assumptions inspired by bottom-up models.
They can therefore better capture domain-specific phenomena,
such as variations in viewing strategies or expectations as to where the most interesting elements are.
They may better capture visual statistical learning, or how people learn to predict where to look on the basis of learned feature distributions \cite{series2013learning}.
However, while deep-learning-based data-driven approaches are currently outperforming stimulus-driven models with natural scenes,
they fall short in terms of ability to capture some pop-out effects in synthetically generated patterns~\cite{He_2019_CVPR}.
Given the strengths and weaknesses of the two approaches,
we decided to investigate both.
In the ``Computational Modeling'' section, we describe the models chosen for our experiments and compare their performance.

\subsubsection{Applications on graphical UIs}

Despite saliency being one of the better-known aspects of the human visual system,
designers often use rules of thumb instead of theory or rely on experimental evaluations \cite{Still10,Rosenholtz11}.
Computational saliency models could be beneficial,
because they can make predictions for the designer about where users are likely to fixate within a given UI.
They can be used, for example,
in delving into the quality of a given design
or as a quick way to facilitate comparisons between designs.
Without necessitating dedicated user studies, they can simulate how a user will attend new, unseen layouts.

Still, the abundant literature on saliency modeling that predicts where humans might look within a scene
contains only a few studies of graphical UIs.
The first computational visual saliency model to predict human attention,
with particular regard to webpages, was proposed by Shen et al.~\cite{Shen14}.
They added several feature maps (text and face detectors, positional bias correction, etc.)
to the classic ITTI saliency model~\cite{Itti98}.
Even with \emph{ad-hoc} training, they could not achieve competitive performance (correlation with ground truth was 0.45).
Xu et al.~\cite{Xu16_chi} proposed a bottom-up approach
that was specifically designed with WIMP graphical user interfaces in mind.
The model takes information about the UI alongside users' mouse and keyboard actions as input
to predict joint spatiotemporal attention maps.
Regrettably, this setup cannot be applied for investigating saliency in mobile UIs, where no keyboard or mouse is available.
Vidyapu et al.~\cite{Vidyapu19} were interested in predicting visual attention \emph{scanpaths} for webpage \emph{images}
and did not investigate other webpage elements, such as headers, navigation menus, and paragraphs of text.
Finally, as mentioned earlier, Gupta et al. \cite{Gupta18} compared stimulus- and data-driven models against crowdsourced mobile eye-tracking data,
concluding that a data-driven deep learning model has the highest fit.
However, they selected the elements in the top 20\% for saliency as their ground-truth dataset,
which makes prediction less challenging (we predict saliency for the whole UI).

\subsection{Visual impression, importance, and clutter}

HCI research has looked also at constructs that come close to saliency but are not quite the same.
\emph{Visual impression} refers to perceived aesthetics of a graphical UI formed during free viewing.
This is typically measured via rating scales,
with results reported for both desktop~\cite{Lindgaard19_web} and mobile interfaces~\cite{Miniukovich14_mobile}.
Visual saliency is a construct related to the control of visual attention, not regarding aesthetics or design quality.
A concept more closely related to saliency is that of \emph{visual importance}.
Bylinskii et al.~\cite{Bylinskii17} extended a pre-trained neural network~\cite{Shelhamer17}
for predicting which regions in a graphic design are felt to be more important.
Importance was measured by means of cursor exploration of a blurred page.
However, a ``poor man's eye-tracker''~\cite{Cooke06},
which involves an element of introspective judgment of importance,
is not a good proxy for gauging visual saliency~\cite{Tavakoli_2017_CVPR}.
Finally, research on \emph{visual clutter} is directly motivated by theories of saliency.
The work of Rosenholtz \cite{Rosenholtz11} shows how models of visual saliency can be exploited to compute indices of how cluttered a display is perceived to be.

\section{Method}
\label{sec:experiments}

\begin{figure*}[!ht]
\centering
\def\w{0.15\linewidth}
\fbox{\includegraphics[width=\w]{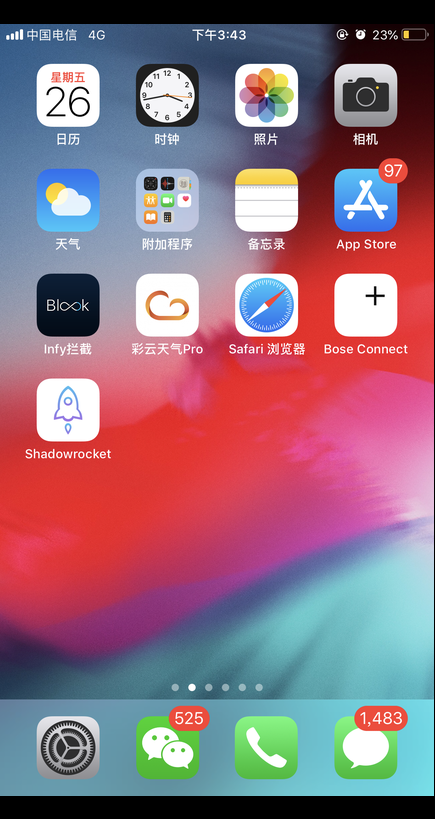}}\hfill
\fbox{\includegraphics[width=\w]{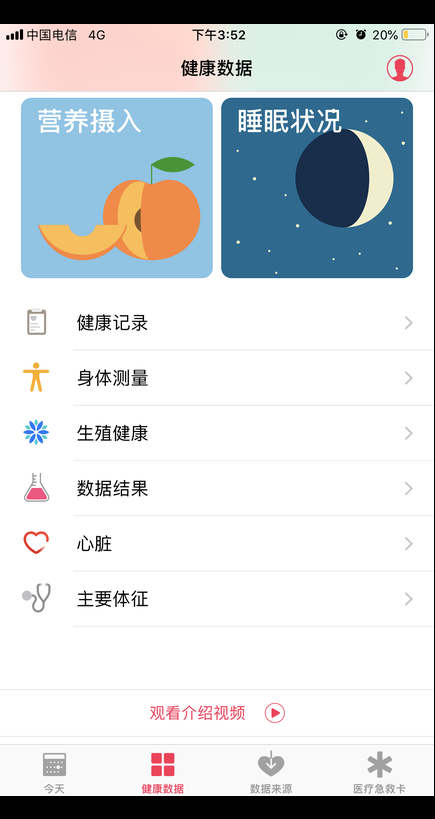}}\hfill
\fbox{\includegraphics[width=\w]{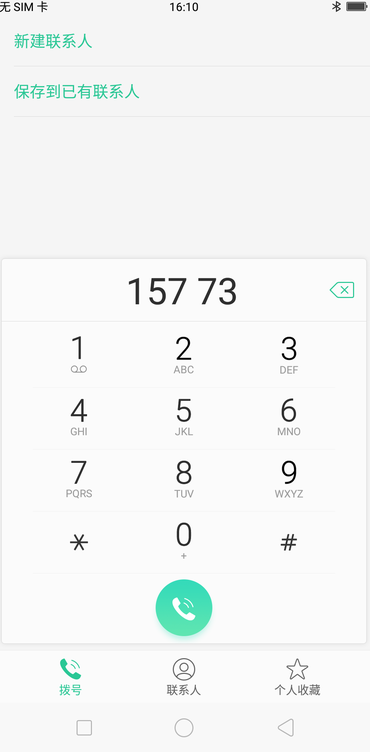}}\hfill
\fbox{\includegraphics[width=\w]{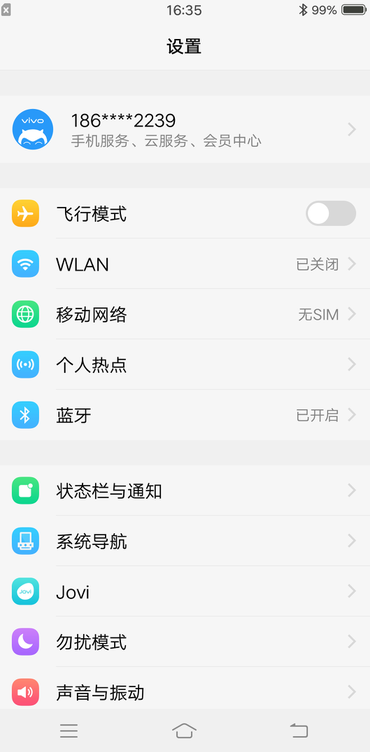}}\hfill
\fbox{\includegraphics[width=\w]{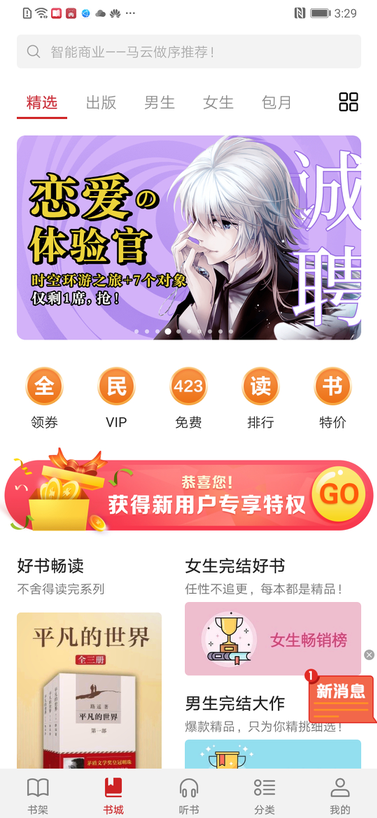}}\hfill
\fbox{\includegraphics[width=\w]{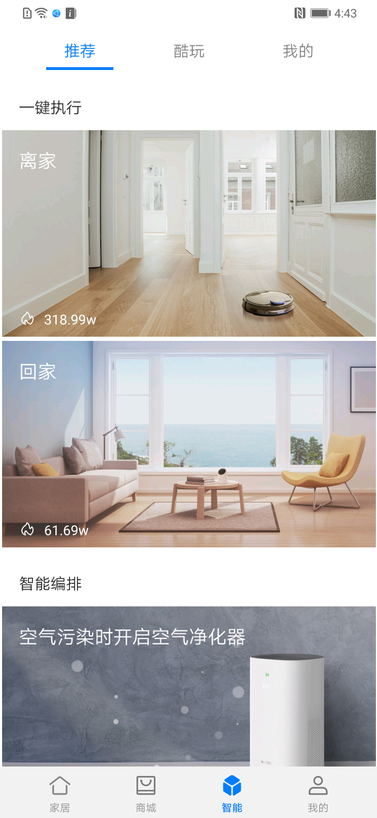}}
\caption{Sample UIs used in the experiment. Our dataset contains a total of 193 screenshots.}
\Description{Six UI screenshots illustrate the kind of apps we analyzed.}
\label{fig:apps}
\end{figure*}

The design of our experiment follows earlier studies on desktop settings~\cite{Shen14},
which we accordingly adapt to our mobile setup~\cite{Long17}.
We collected a large set of representative mobile interfaces,
which in the study were viewed from a distance resembling that of mobile use.
The set of mobile UIs is varied and has a rich distribution of different element types.
We used a carefully calibrated high-fidelity eye tracker to collect gaze data during free viewing.
Note that an alternative setup, for example including walking (e.g., on a treadmill or a closed circuit)
would mainly add nuisance factors that would detract from the quality of data and would require significantly larger sample sizes.

\subsection{Participants}

Thirty participants (12 male, 18 female) were recruited via mailing lists and announcements in social media.
The average age was 25.9 (SD=3.95).
The participants had normal vision (8) or corrected-to-normal-vision (22).
Twenty of the 22 wore glasses and the remaining two wore contact lenses.
No participant was color blind.
All participants were native Chinese speakers,
the language used in the mobile interface dataset (see below).

\subsection{Materials}

We collected a dataset comprising 193 UI screenshots of different mobile apps
from Android and iOS devices (vendors: Apple, Huawei, Oppo, and Vivo).
The screenshots were taken from mobile stock applications, available via application markets.
All screenshots were shown in portrait mode,
since all apps were designed to be operated in portrait orientation,
and scrolled up, to mimic what any user would see upon launching any app.
Particular care was taken to cover a wide range of commonly used UIs, including e.g.
home screen, settings, gallery, camera, contacts, music, recorder, calendar, calculator, notepad, file explorer.
All UIs used simplified Chinese language, though they are also available in other languages.

\autoref{fig:apps} illustrates some of these screenshots.
More examples are given in the \emph{Supplementary Materials}.
The screenshots were taken on different high-definition smartphones,
and thus had different screen resolutions:
1242x2208 (Apple iPhone 8 plus), 1080x2340 (Huawei P30), and 1080x2160 (Oppo R11s, Vivo X20).
Upon presentation in the experiment, they were resized to the lowest common resolution (1080x2160),
scaled down 35\% so that they could fit in a Full HD monitor (1920x1080),
and centered on the screen which had a black background.
No distortion artifacts appeared after resizing.
This allowed us to display all UIs to the users
regardless the actual contrasts and pixels densities (e.g. Retina display on the iPhone)
of the different smartphones.
\autoref{fig:setting} shows one of these screenshots used as experimental stimuli.

\subsubsection{Element-level segmentation}
We also manually segmented and labeled UI elements on the screenshots to form an element taxonomy.
For this, we followed existing design guidelines
and previous research~\cite{Apple:guidelines, Google:guidelines, Liu:2018:RICO, MobileUI:guidelines, SemanticUI:guidelines}.
The taxonomy was determined by two human coders, through a consensus-driven, iterative process.
Several commonly-used, human-perceptual level UI element categories were identified,
refined, re-categorized, and finally reviewed by academic researchers and industry experts.
The final version is summarized in \autoref{tbl:dataset:categories} and more thoroughly in the \emph{Supplementary Materials}.
The fine-grained semantic information on elements permits detailed analyses of how UI elements may affect visual saliency.

\begin{table}[b!]
\centering
\begin{tabular}{|l|l|l|l|l|l|l|l|}
\hline
Icon &
Header &
Text &
Text button &
Text group &
Label &
Image &
Card \\
\hline
Popup &
Dialog box &
Multi-tab &
Bottom bar &
Switch &
Slider &
Date picker &
Search bar \\
\hline
\end{tabular}
\caption{Taxonomy used for element-level segmentation of UI elements in our mobile UI dataset.
  See the Supplementary Materials for examples of each element type.
}
\label{tbl:dataset:categories}
\end{table}

\subsection{Task and Experimental Design}

The task was free-viewing a set of mobile UI screenshots.
Each UI is shown for 3.5\,seconds, informed by previous work \cite{Shen14}
that noticed that fixations start to diverge after 3\,s.
Then there is a blackout period of 1.5\,s
before auto-advancing to the next trail,
which we found to be an adequate tradeoff while avoiding visual fatigue.
The order of the stimulus UIs was randomized per participant.

\subsection{Apparatus}

We used the SMI RED250 eye-tracker,
which has a sampling rate of 250\,Hz and an accuracy of 0.4\,degrees.
A stationary eye-tracker ensures very high quality of data,
impossible to achieve presently with mobile trackers.
We used the SMI Experiment Center software to implement the experimental design,
calibrate the eye-tracker to each participant,
compute the fixation data,
and automate the administration of visual stimuli.
Then we used SMI BeGaze software to save and export the recorded data.
Besides fixations data, the software generates gaze paths and heatmaps.
The desktop monitor was 24.5'' wide and had full HD resolution (1920x1080).

\subsection{Setup}

\begin{figure}[!ht]
\centering
\includegraphics[width=0.8\linewidth]{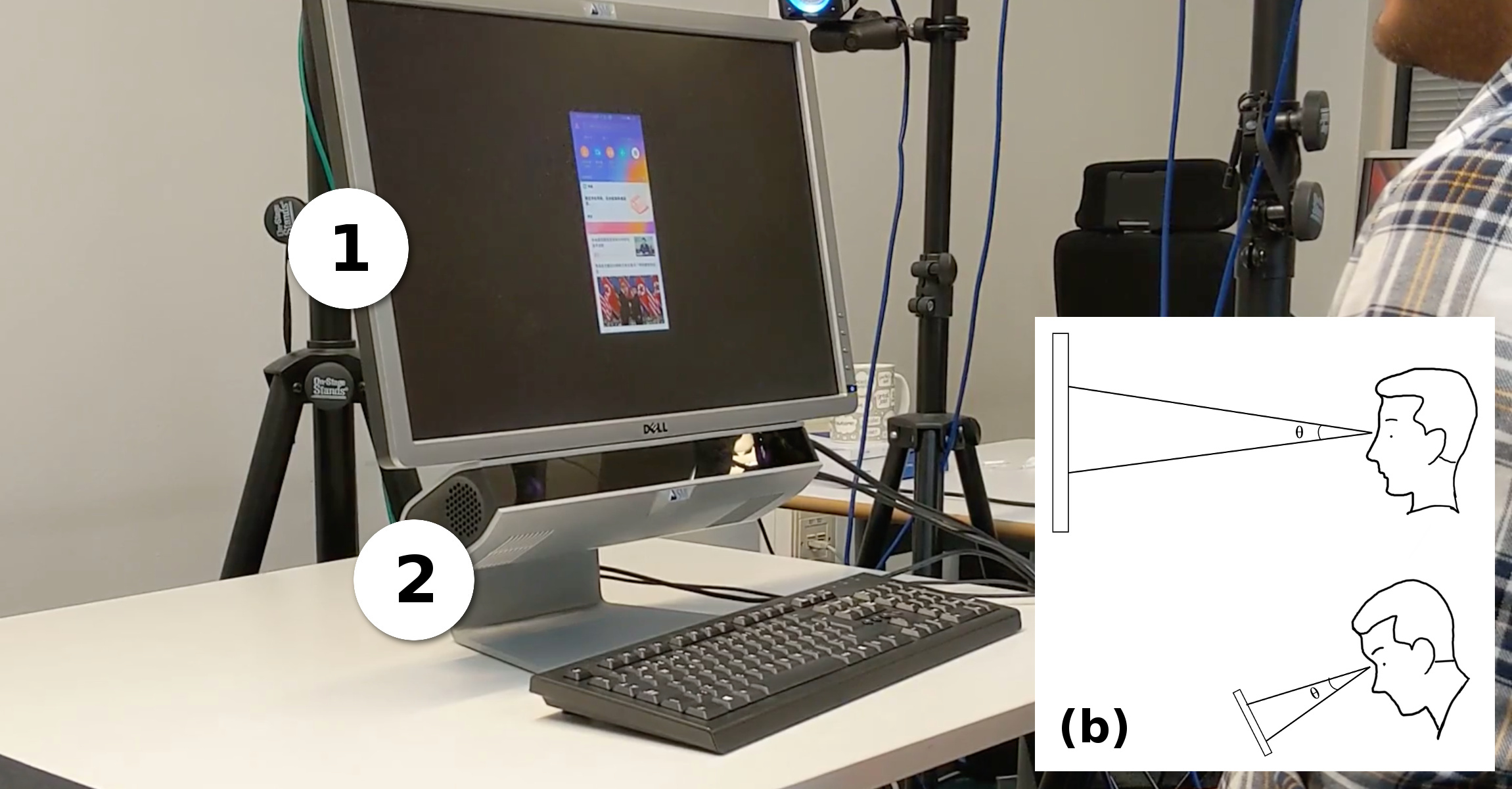}
\caption{
  Experimental setup, depicting (1)~the monitor where the UIs were displayed and (2)~the eye-tracker.
  The keyboard was only used by the experimenter.
  The inner diagram~(b) illustrates how our setting preserved the viewing angle of a mobile scenario.
}
\Description{A man is looking at a monitor with a UI screenshot in the middle of the screen.}
\label{fig:setting}
\end{figure}

Participants sat comfortably in front of the eye-tracker, see \autoref{fig:setting}.
The height of the desk was adjusted to the height that best suited each individual.
We advised participants not to lean toward the monitor.
We corrected for screen distance and the size of screenshot stimuli according to Long et al.~\cite{Long17},
who reported that the mean viewing distance to a mobile phone is 29.2 $\pm$ 7.3\,cm.
Then, considering that the physical size of the displayed stimuli is roughly twice larger than it is on a mobile screen,
the distance between participants' eyes and the screen was kept within the 60--70\,cm range in our experiment.
Therefore, our setup approximates well the defining real-world parameters affecting the data.

\subsection{Procedure}

At the beginning of the experiment, participants were told about the goals of the study
and were provided with informed consent.
The eye-tracker was calibrated for each participant using the usual 3x3 point grid.

Each participant was exposed up to three batches of 60 mobile UI screenshots each, to avoid visual fatigue.
They were told to look at each UI screenshot without any prescribed aim (free viewing task),
since asking users to search for a particular goal would be affected not only by saliency
but also by expectations, location memory, search strategies, etc.
As previously hinted, screenshots were shown in randomized order for 3.5\,seconds, followed by 1.5\,seconds of blank screen.
The software automatically advanced to the next UI screenshot, until the end of the batch.
After this, participants could take a rest as long as they needed
and then were asked to either proceed with the next batch or finish the experiment.
Nineteen users completed the three batches,
ten users completed two batches, and one user did only one batch.
A batch is 5\,min long and the average session lasted 13\,min.
Each UI was assessed by 24 participants on average (min 16, max 29).

To help the participants get familiarized with the experimental procedure,
a warm-up session was always conducted before starting the actual data collection.
Participants were allowed to talk with and ask questions to the experimenter during the warm-up session.
Four UI screenshots were used in the warm-up session, which were not shown in the actual experiment.

\subsection{Data preprocessing}
Fixations that happened outside the viewport of the mobile UI (8\% of the fixations)
were not included in the following analyses.
The dataset we release includes the raw data.

\section{Results}
\label{sec:results-empirical}
The data permit revisiting the phenomena found in studies with natural scenes, discussed above.

\subsection{Effect of location}

\begin{figure}[!ht]
\centering
\subfigure[Fixation location bias\label{fig:bias:location}]{
    \includegraphics[width=0.5\linewidth]{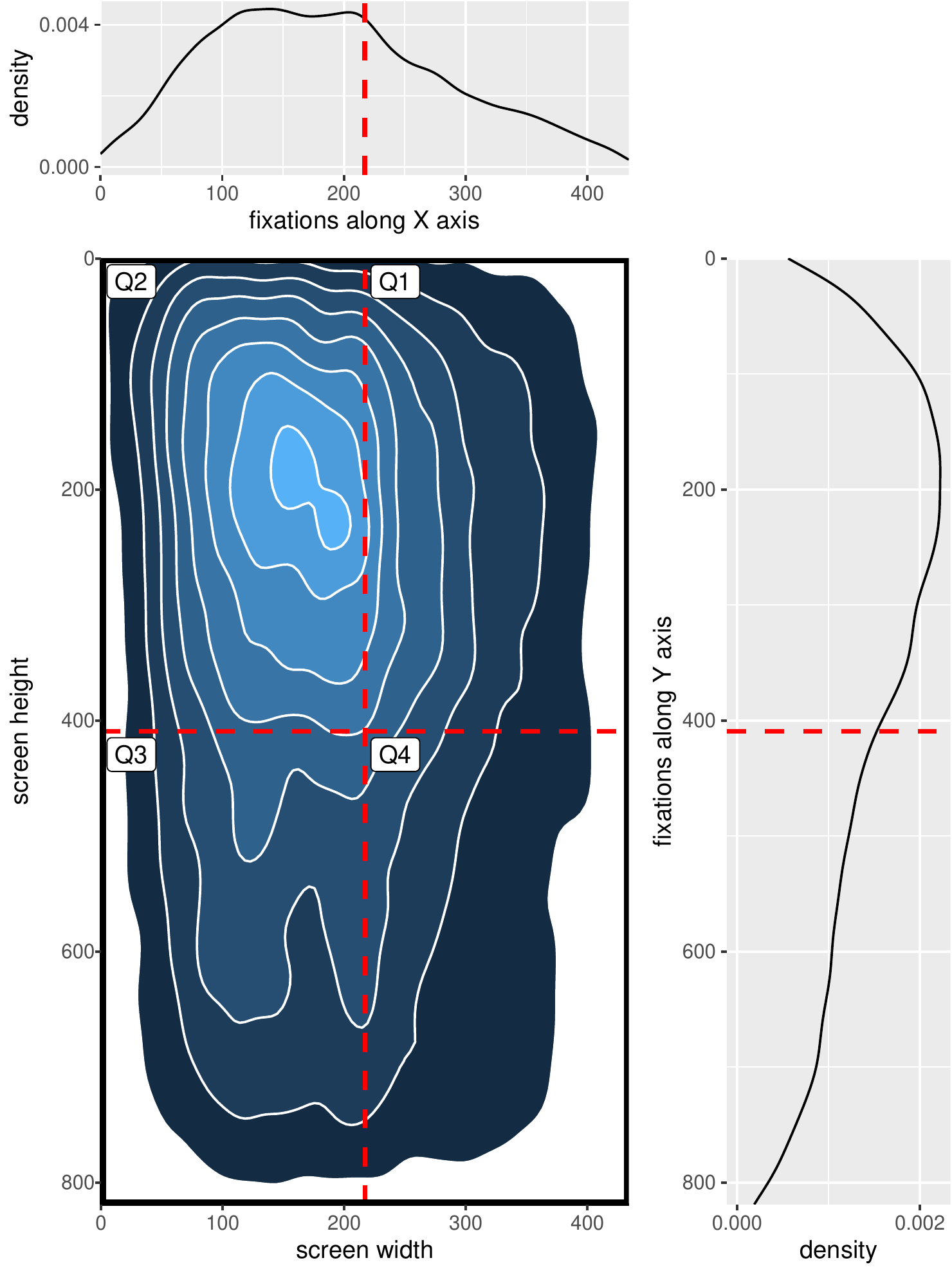}}
\hfill
\subfigure[Color brightness bias\label{fig:bias:color}]{
    \includegraphics[width=0.35\linewidth]{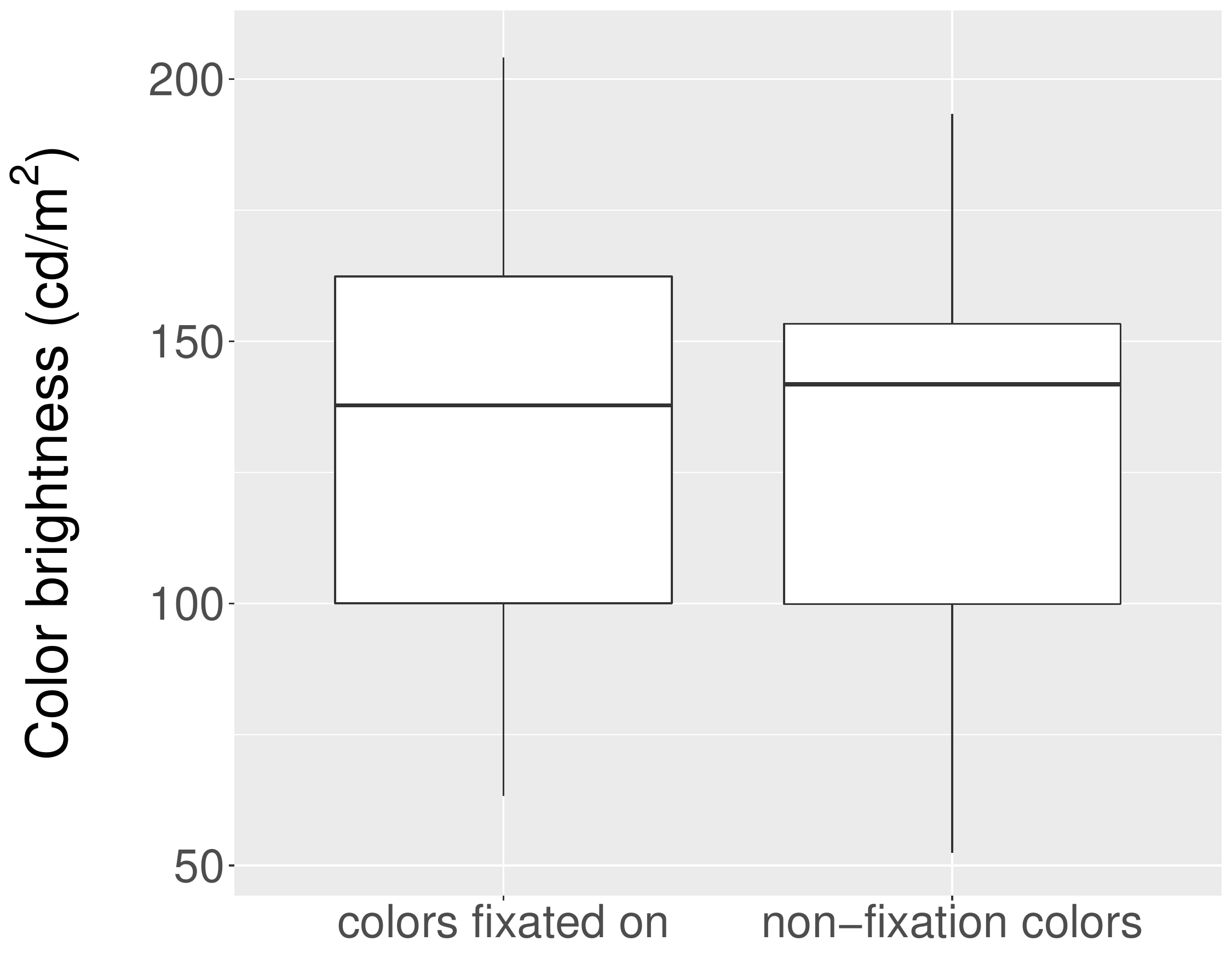}}
\caption{Effects of fixation location (a) and color brightness (b) on saliency.
    One can see that fixations happen often in brighter colors and are highly skewed toward Q2;
    i.e., there is no clear horizontal or center bias.
    The thick dashed lines (a) denote the middle of the mobile screen along the X and Y axis.}
\Description{A plot of fixation distributions is shown at the left. A breakdown of such fixations on the basis of color brightness is shown at the right.}
\end{figure}

We observed a strong location bias.
\autoref{fig:bias:location} shows the spatial distribution of eye fixations across the quadrants of the mobile display.
The top-left quadrant (Q2) attracted the most fixations (43.3\%).
Quadrants 1 (top right) and 3 (bottom left) together attracted about as many as Q2.
Quadrant 4 showed clearly the fewest fixations (13.1\%).

The omnibus test revealed a statistically significant difference
between the average number of fixations per user and UI for each of the quadrants:
$\chi^2_{(3, N=1347)} = 357.65, p < .001, \phi = 0.515$.
Effect size $\phi$ suggests a high practical importance of the differences~\cite{Kim17}.
We then ran Bonferroni-Holm corrected pairwise comparisons as post-hoc test
and found that Q2 attracted significantly more fixations than any other part of the UI,
followed by the top right (Q1), bottom left (Q3), and bottom right (Q4).
No statistically significant difference was found between Q1 and Q3 ($p > .05$).
All other comparison results were found to be statistically significant ($p < .001$).

\subsubsection{Summary}
Had there been a horizontal bias, we would have expected to see Q2 vs. Q1 and/or Q3 vs. Q4
differing markedly in their proportion of fixations.
If there had been a center bias, we should have seen a more equal distribution among the quadrants.
Given the observations made, we conclude that, in contrast, the data suggest a strong top-left location bias.

\subsection{Effect of colors}

\begin{figure}[!ht]
\centering
\subfigure[Color distribution of the pixels displayed \label{fig:bias:colors:original}]{
  \includegraphics[trim=50 140 30 110, clip, width=0.49\linewidth]{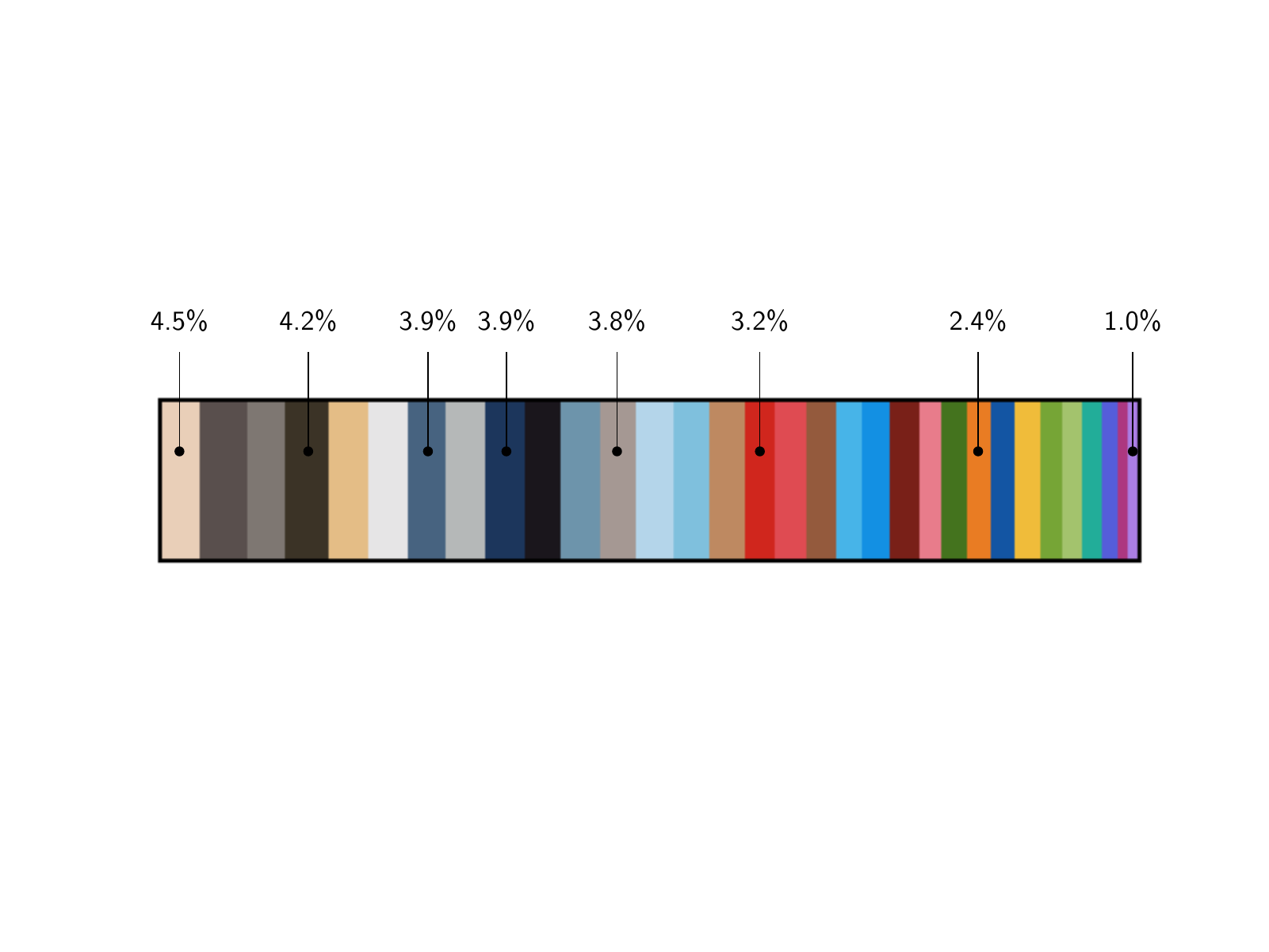}}
\hfill
\subfigure[Color distribution of the pixels fixated on \label{fig:bias:colors:fixations}]{
  \includegraphics[trim=50 140 30 110, clip, width=0.49\linewidth]{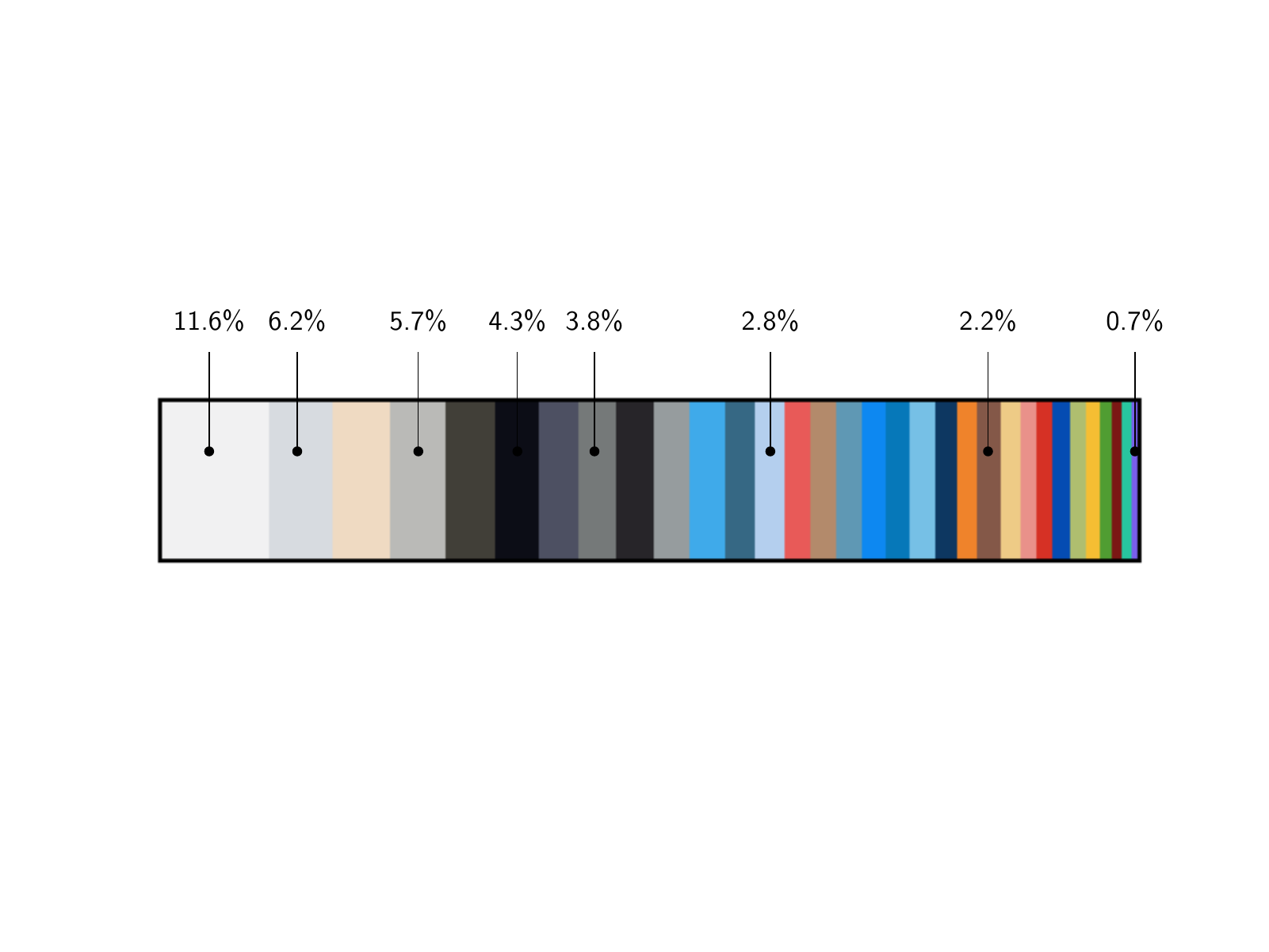}}
\caption{Colors displayed vs. those fixated upon, sorted by frequency.
  Although brighter colors tend to receive more attention,
  we found no significant difference.}
\Description{A plot of the distribution of fixated colors is shown at the left. A plot of the distribution of non-fixated colors is shown at the right.}
\label{fig:bias:colors32}
\end{figure}

We did not observe an effect of color brightness.
We started by checking whether the colors presented and those fixated on show the same distribution.
\autoref{fig:bias:colors32} shows the 32 most prevalent colors in the original UIs and the top 32 colors by number of fixations.
Comparison suggests that brighter colors may attract more attention than darker colors.

To investigate whether a reliable effect exists,
we computed the pixel brightness values by using sRGB Luma coefficients (ITU Rec.\,709)~\cite{Bezryadin07},
which reflect the corresponding standard chromaticities,
and compared the distribution of fixation and non-fixation brightness values.
\autoref{fig:bias:color} shows the data.
Bartlett's test of homogeneity of variances showed non-significance ($T = 0.392, p = .531, d = 0.36$),
suggesting that brighter colors do not attract significantly more fixations than darker ones.

\subsubsection{Summary}
We did not find a color bias affecting saliency of mobile UIs.
If attention is drawn to brighter colors, the effect is too modest to be of practical significance.

\subsection{Effect of element type}

\begin{figure*}[!ht]
\centering
\def\w{0.24\textwidth}
\includegraphics[width=\w]{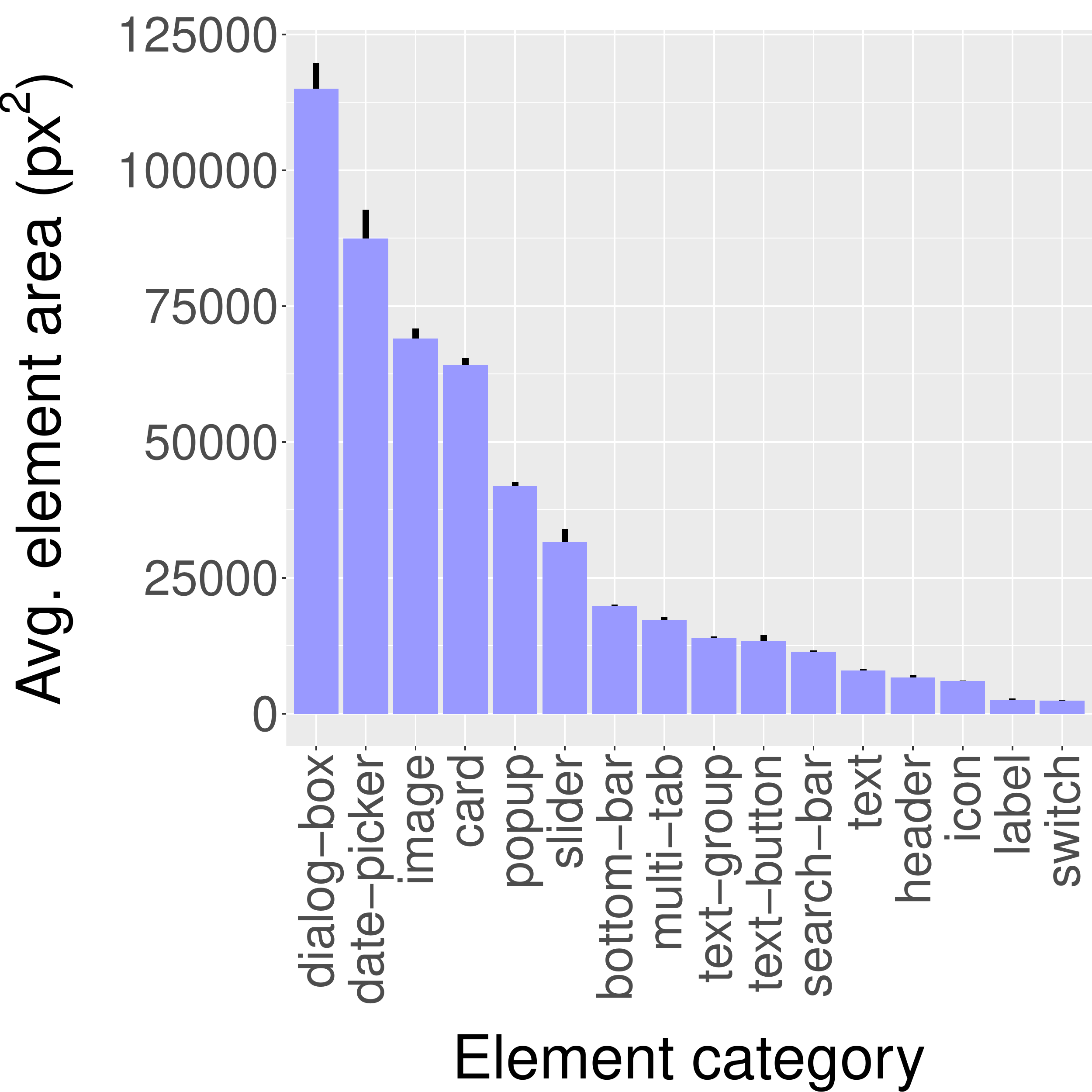}\hfill
\includegraphics[width=\w]{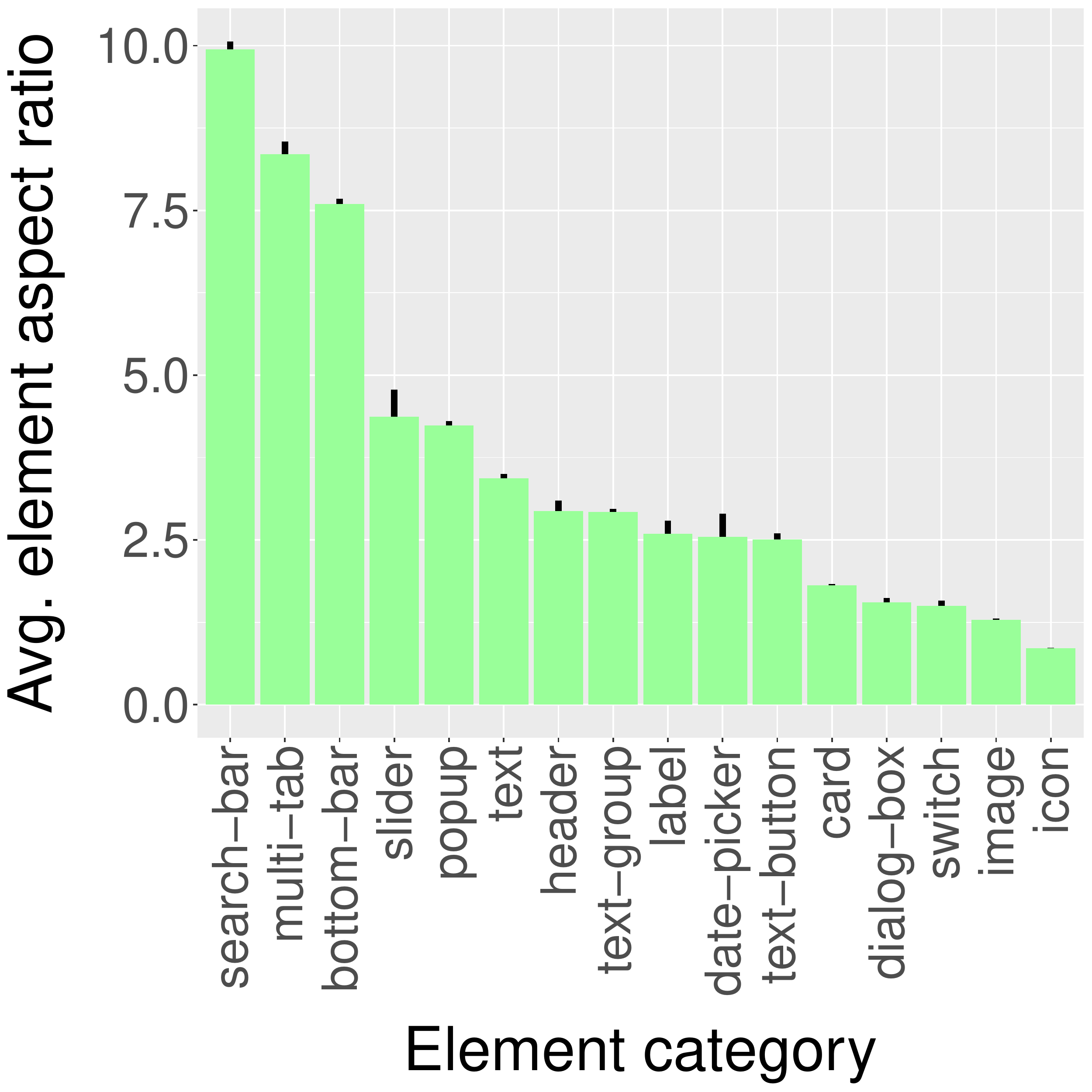}\hfill
\includegraphics[width=\w]{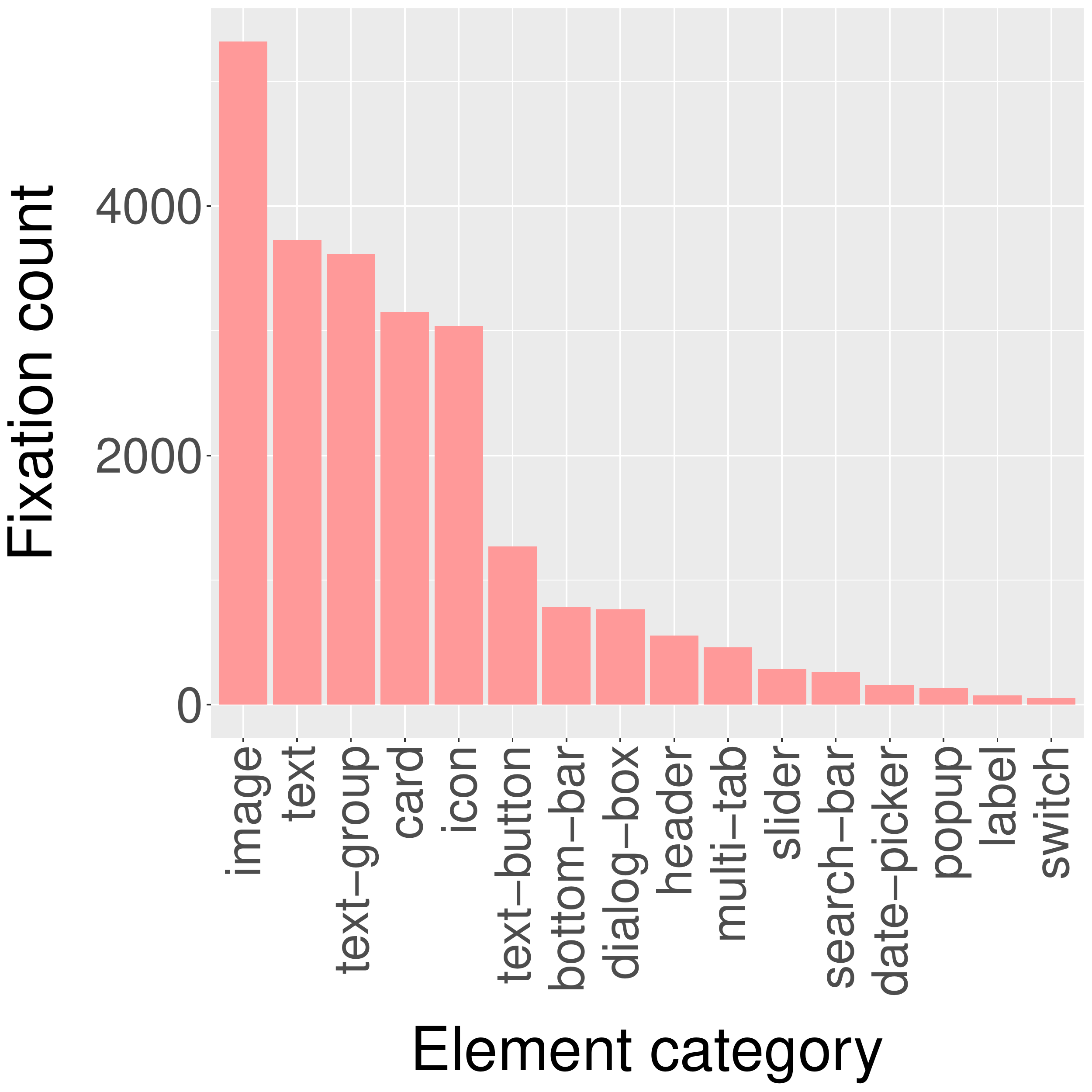}\hfill
\includegraphics[width=\w]{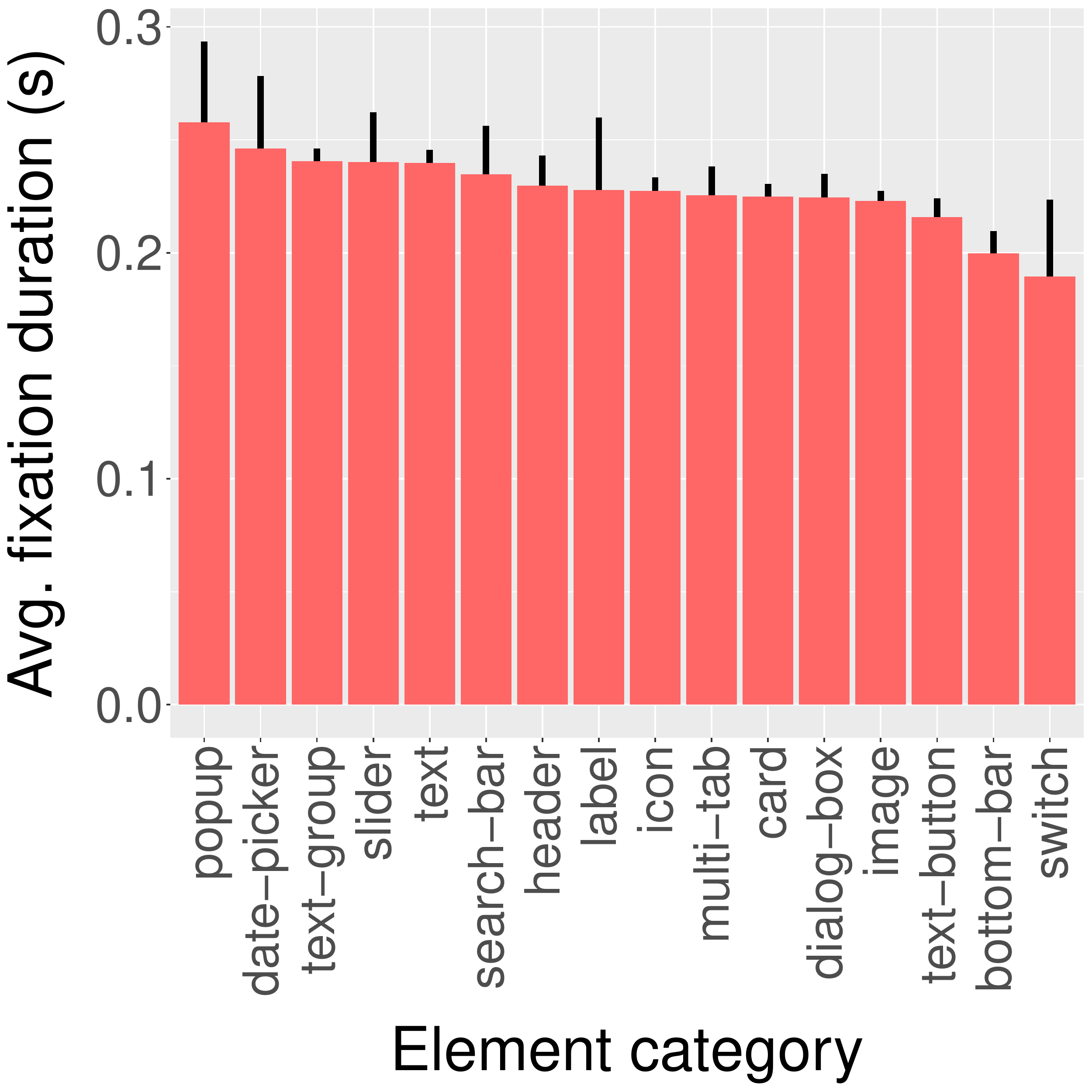}
\caption{Base rates plotted against fixations, by element type.
Error bars denote 95\% confidence intervals.}
\Description{Four plots are shown to illustrate the effect of element types on saliency according to: area size, aspect ratio, number of fixations, and fixation duration.}
\label{fig:bias:elements}
\end{figure*}

\textbf{Text:}
We did find bias toward particular UI elements.
Firstly, we saw that most fixations recorded (58.56\%) were on the annotated UI elements.
Most of the remaining fixations were on blank areas of the UI,
such as the space between icons and text.
\autoref{fig:bias:elements} depicts the distribution of fixation frequency,
fixation duration, element areas, and element aspect ratios.
As can be observed,
while all UI element categories attracted approximately the same fixation duration on average
(between 0.2 and 0.3 seconds),
five element categories received clearly more fixations than the rest.
The UI elements fixated upon most often were
``Image'' (22.4\%), ``Text'' (15.7\%), ``Text group'' (15.2\%), ``Card'' (13.3\%), and ``Icon'' (12.8\%).
The remaining UI element categories attracted below 10\% of the total fixations each.
Combined, the two text categories account for 30.9\% of all fixations on UI elements.
Therefore, we confirmed the text bias to be present also in mobile UIs.

\textbf{Images:} Images, similarly, received more attention than was expected from the area they cover.
However, images are more varied in content.
We manually inspected all UI elements belonging to the ``Image'' or ``Card'' category,
to confirm or deny the existence of the face bias discussed earlier in the paper.
There were 34 UIs in which one or more faces appeared,
thus representing 17.6\% of the UIs in the dataset.
Interestingly, whenever an image or card had only a human face, as was the case in 19 UIs,
fixations occurred at the precise location of the face,
Then, as long as the image had text too (seen with 7 UIs),
the fixations were much closer to the text's location than to the face location.
This effect is illustrated in the bottom portion of \autoref{fig:good-results} (see the ``Ground-truth'' column),
and was found to be statistically significant:
$\chi^2_{(2, N=34)} = 27.137, p < .001, \phi = 0.893$.
Therefore, we confirmed the face bias to be present also in mobile UIs.

\textbf{Size:} Finally, we also investigated whether the size of a UI element had an effect on the fixations made.
We ran Pearson's product-moment correlation tests between area vs. number of fixations made ($r = 0.08$)
and vs. fixation duration ($r = 0.21$),
and on aspect ratio vs. number of fixations  ($r = -0.39$)
and vs. fixation duration ($r = 0.05$).
No significant correlation was found for any of these ($p > .05$).
Therefore, we conclude that element size does not play an important role in visual saliency with mobile UIs.

\subsubsection{Summary}
We found evidence for a strong text bias and a bias toward images and faces.
There was no evidence for a size bias.

\section{Computational Modeling}
\label{sec:modelling}
In this section,
we assess a number of classic and data-driven computational models
in light of the saliency data collected.
As examples of classic models we consider three well-known stimulus-driven models
that are often used as baseline models in the literature:
ITTI~\cite{Itti98}, GBVS~\cite{Harel07}, and BMS~\cite{Zhang13}.
ITTI employs difference of Gaussians for feature extraction and pools the features together in order to infer saliency.
GBVS uses image pyramids for feature extraction and employs graph diffusion to find salient points.
BMS uses thresholding to obtain feature maps at different scales and combines them to one final map.
Regarding the data-driven deep learning (DL) models, we consider SAM~\cite{Cornia18},
a state-of-the-art model that has publicly available source code,
and ResNet-Sal, a custom DL model we developed on the basis of SAM that has a simpler decoding pipeline.

Our model evaluation used the gaze recordings as ground-truth data.
Since the classic models are training-free, they can be evaluated directly, with no added effort.
The DL models, however, require a large volume of training data.
We explored various ways to train these models,
aiming for the best possible conditions.
Here, we refer to the SALICON dataset~\cite{Jiang2015} (2015 and 2017 releases),
a reference dataset in visual saliency experiments that consists of natural images.
We also fine-tuned SAM to the mobile UI dataset via transfer learning~\cite{Goodfellow16},
so that the model could learn particular characteristics of our graphical interfaces.
Note that no model uses the UI annotations as an additional feature
because they are designed to predict bottom-up saliency.

\subsection{Model implementations}

The three classic computational models have public implementations in Matlab code,
while the DL models are available in Python.
In SAM, the popular ResNet-50 convolutional neural network~\cite{He16} is used
as a pre-trained feature-encoding backbone,
and an attentive ConvLSTM recurrent neural network acts as an integration (saliency decoding) network.
To better understand the role of such a sophisticated attentive decoding network,
we developed ResNet-Sal, which uses the same encoding architecture as SAM
and a non-attentive, much simpler decoding network
composed of three upscaling blocks (each having a convolution transpose)
followed by two 2D convolution layers.

In summary, we used three classic saliency models ``as-is'' and the following DL models:
\begin{description}
    \item[SAM-S2015:] Encoding network pre-trained on ResNet-50 weights.
        Attentive decoding network trained on SALICON 2015 dataset.
    \item[SAM-S2017:] Encoding network pre-trained on ResNet-50 weights.
        Attentive decoding network trained on SALICON 2017 dataset.
    \item[SAM-mobile:] Encoding network pre-trained on ResNet-50 weights.
        Attentive decoding network trained on SALICON 2017 dataset
        and fine-tuned to our mobile dataset.
    \item[ResNet-Sal:] Encoding network pre-trained on ResNet-50 weights.
        Non-attentive decoding network trained on SALICON 2017 dataset
        and fine-tuned to our mobile dataset.
\end{description}
As noted, all DL models have the same pre-trained encoding network
but the SAM variants use an attentive decoding network trained on different datasets.

\subsection{Training and testing}

Our dataset is divided into a training partition, comprising 80\% of the screenshots and their associated fixation data,
and a test partition, consisting of the remaining 20\% of the data.
We took special care to balance the number of representative screenshots from each mobile vendor
in both the training and the test partition.
We used the RMSprop optimizer~\cite{tieleman2012lecture}
and the following loss function~\cite{Cornia18}:
\begin{equation}\label{eq:loss}
\mathcal{L}(S,F) = \alpha \, \text{NSS}(S,F^B) + \beta \, \text{CC}(S,F^C) + \gamma \, \text{KL}(S,F^C)
\end{equation}
where $S$ is the predicted saliency map and
$F$ is the ground-truth fixation data, in either continuous ($F^C$) or binary ($F^B$) form.
As described below, NSS is the normalized scanpath saliency,
CC is the correlation coefficient, and KL refers to the Kullback-Leibler divergence,
which are commonly used to evaluate saliency prediction models.
As observed, the model places particular emphasis on predicting a saliency distribution
that matches the ground-truth human visual saliency as closely as possible.
The loss scalars above are set to $\alpha=-1$, $\beta=-2$, $\gamma=10$,
as suggested by \citet{Cornia18}.

\subsection{Ground truth}

The eye tracker provides fixation points ($F^B$ in the equations above)
so we computed the ground-truth continuous density map $F^C$
by smoothing each fixation point with a 2D Gaussian filter using a standard deviation of 25\,px.
This size approximates that of the foveal region of the human eye~\cite{Shen14},
as 1 visual degree approximates 50\,px in our experimental setup.
As stated previously, fixation points falling outside the viewport of the mobile UIs
were not taken into account.

\subsection{Evaluation metrics}

Given the goal of predicting the fixation locations in an image,
a saliency map can be interpreted as a classifier of which pixels do or do not receive fixations~\cite{Bylinskii19}.
With that notion in mind, the literature has proposed the following set of evaluation metrics
to benchmark the performance of a given saliency model.
Some metrics have been designed specifically for saliency evaluation (NSS),
while others have been adapted from signal detection (AUC),
image matching and retrieval (SIM),
and statistics (CC).
We report results on all of them.

\emph{Area under the ROC curve (AUC):}
There are several variants of the AUC location-based metric,
from among which we chose Judd et al.'s~\cite{Judd12}
for historical reasons -- for example, the popular MIT benchmark ranks models based on the AUC-Judd score~\cite{Borji19}.
The saliency map is treated as a binary classifier of fixations for various threshold values (level sets),
and an ROC curve (true-positive rate vs. false-positive rate) is swept out
by measuring the true- and false-positive rates under each binary classifier.
Hence, the higher the AUC value, the better,
indicating a greater ability to predict the salient locations in an image.

\emph{Normalized scanpath saliency (NSS):}
The NSS location-based metric is computed as the average normalized saliency at fixation locations.
Unlike with AUC, the absolute saliency values are part of the normalization calculation.
Given a saliency map $S$ and a binary map of fixation locations $F^B$,
\begin{equation}\label{eq:nss}
\text{NSS}(S, F^B) = \frac{1}{N} \sum_i \bar{S}_i \, F^B_i
\end{equation}
where $N = \sum_i F^B_i$ and $\bar{S} = \frac{S - \mu(S)}{\sigma(S)}$.
A higher NSS value is better,
indicating that the predicted map accumulates more saliency at the fixation points.

\emph{Similarity (SIM):}
SIM is a distribution-based metric for the similarity between two map distributions, viewed as histograms.
It is computed as the sum of the minimum values at each pixel, after normalization of the input maps.
Given a saliency map $S$ and a continuous map of fixation locations $F^C$,
\begin{equation}\label{eq:sim}
\text{SIM}(S, F^C) = \sum_i \min( S_i, F^C_i )
\end{equation}
where $\sum_i S_i = \sum_i F^C_i = 1$.
The higher the SIM value, the better,
indicating greater consistency with human saliency.

\emph{Correlation coefficient (CC):}
Pearson's Correlation Coefficient is a distribution-based metric that measures how correlated or dependent two map distributions are:
\begin{equation}\label{eq:cc}
\text{CC}(S, F^C) = \frac{\sigma(S, F^C)}{\sigma(S) \, \sigma(F^C)}
\end{equation}
where $\sigma(S, F^C)$ is the covariance of $S$ and $F^C$.
The higher the CC value, the better,
though CC is a bounded score and CC=1 denotes the predicted saliency map perfectly matching the ground-truth distribution.
In other words, higher CC values indicate greater consistency with human saliency.

\subsection{Results}
\label{sec:results-modeling}
\begin{figure*}[!htpb]
  \centering
  \setlength{\tabcolsep}{2pt}
  \def\w{0.104\linewidth}

  \begin{tabular}{*9c}

    \tth{Source UI} & \tth{Ground-truth}
    & \tth{SAM-mobile} & \tth{SAM-S2015} & \tth{SAM-S2017} & \tth{ResNet-Sal} & \tth{GBVS} & \tth{BMS} & \tth{ITTI} \\

    \includegraphics[width=\w]{figs/saliency/vivo1.png} &
    \includegraphics[width=\w]{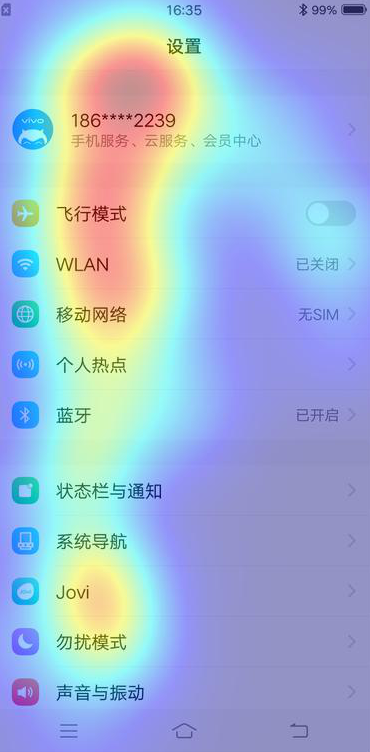} &
    \includegraphics[width=\w]{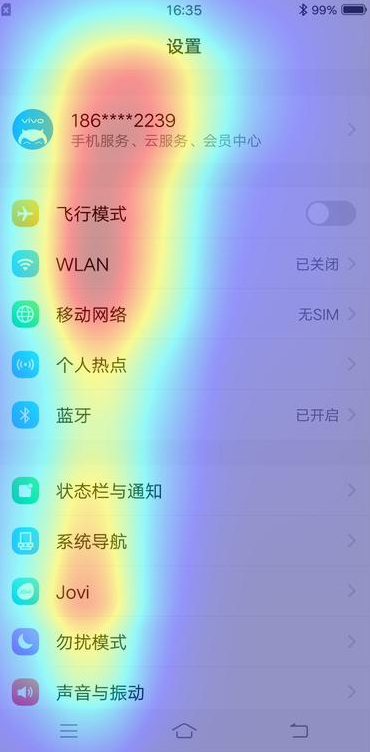} &
    \includegraphics[width=\w]{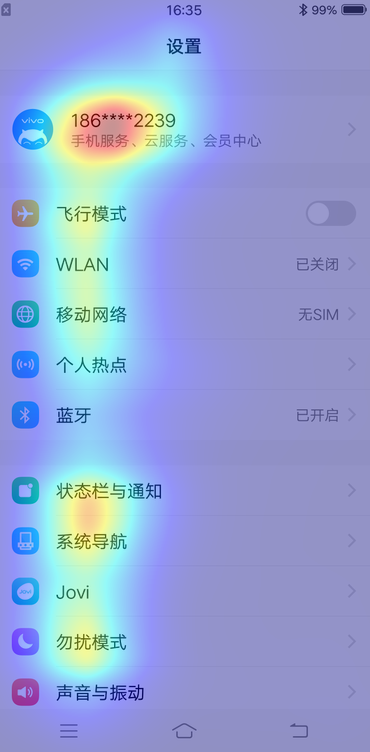} &
    \includegraphics[width=\w]{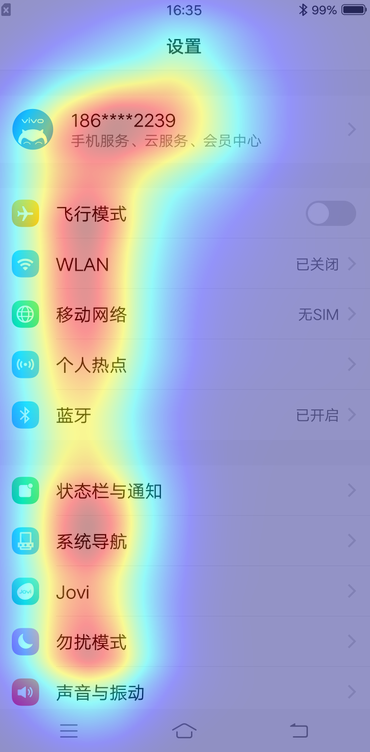} &
    \includegraphics[width=\w]{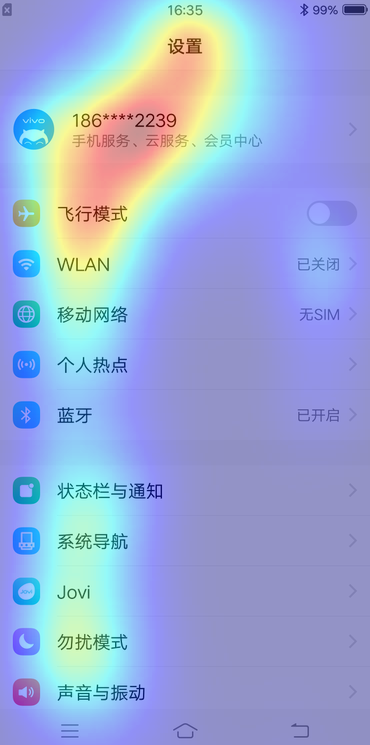} &
    \includegraphics[width=\w]{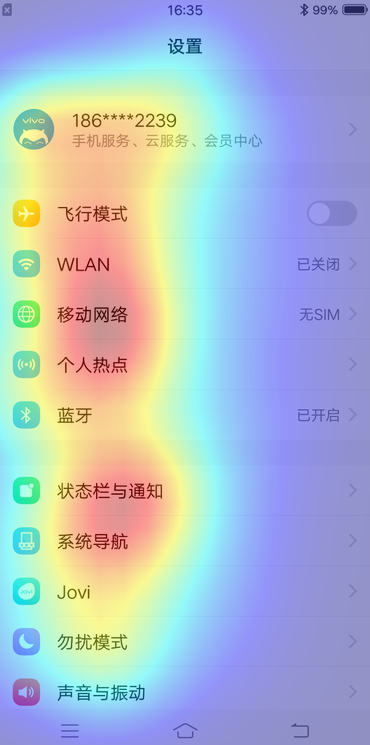} &
    \includegraphics[width=\w]{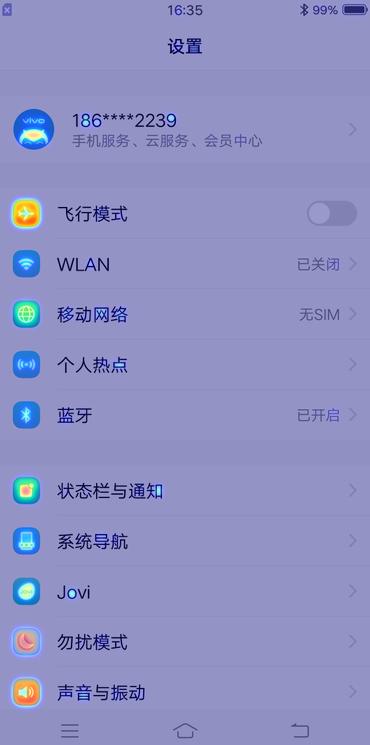} &
    \includegraphics[width=\w]{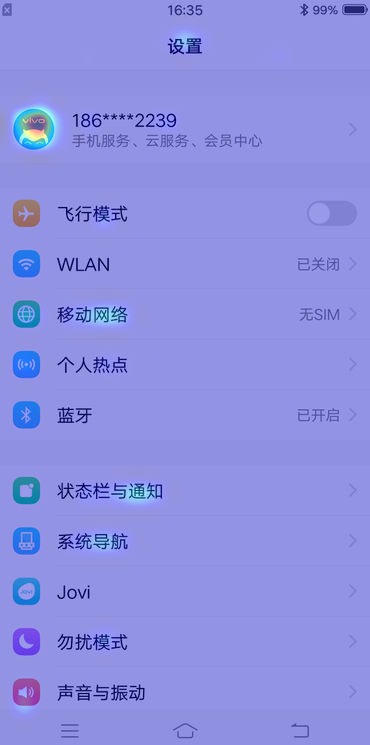} \\

    \tth{Source UI} & \tth{Ground-truth}
    & \tth{SAM-mobile} & \tth{SAM-S2015} & \tth{SAM-S2017} & \tth{ResNet-Sal} & \tth{GBVS} & \tth{BMS} & \tth{ITTI} \\

    \includegraphics[width=\w]{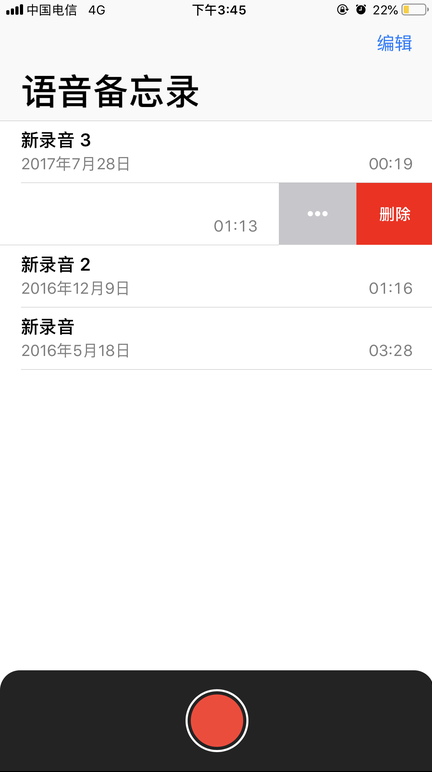} &
    \includegraphics[width=\w]{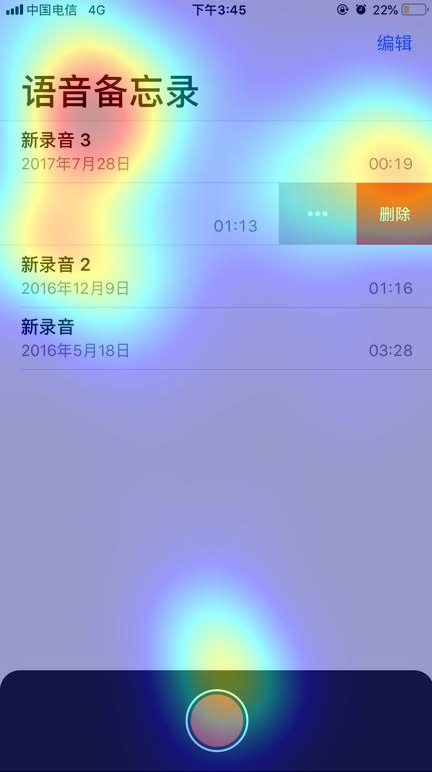} &
    \includegraphics[width=\w]{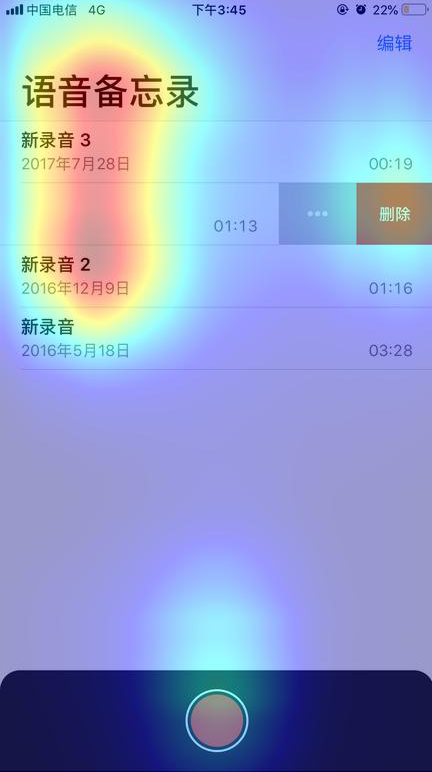} &
    \includegraphics[width=\w]{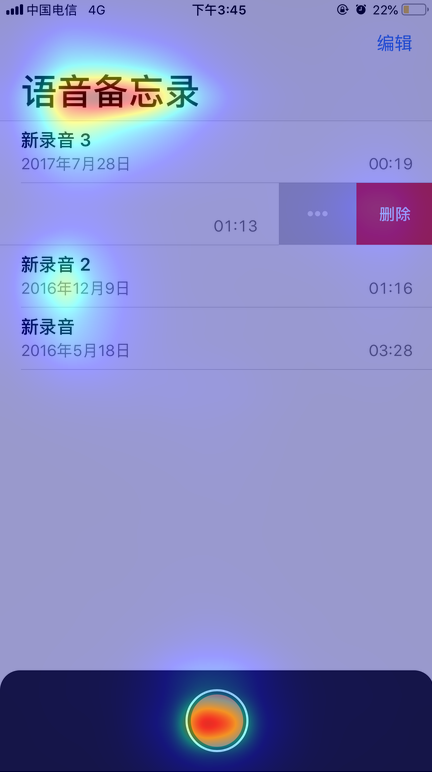} &
    \includegraphics[width=\w]{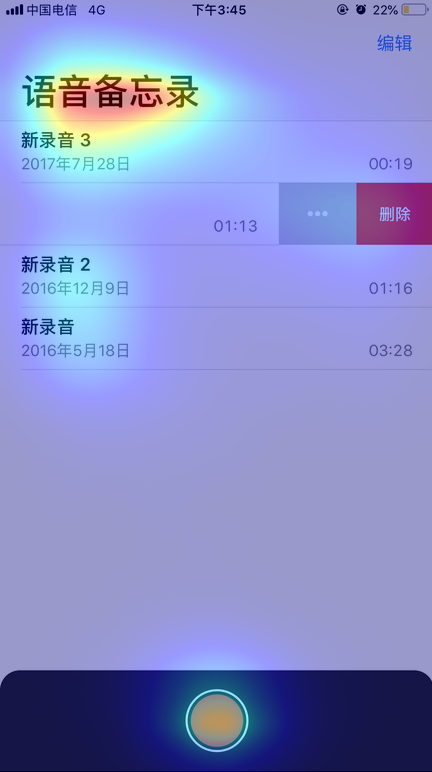} &
    \includegraphics[width=\w]{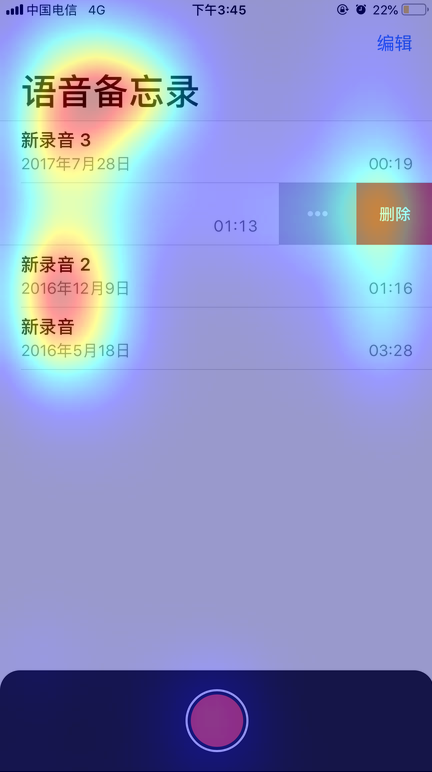} &
    \includegraphics[width=\w]{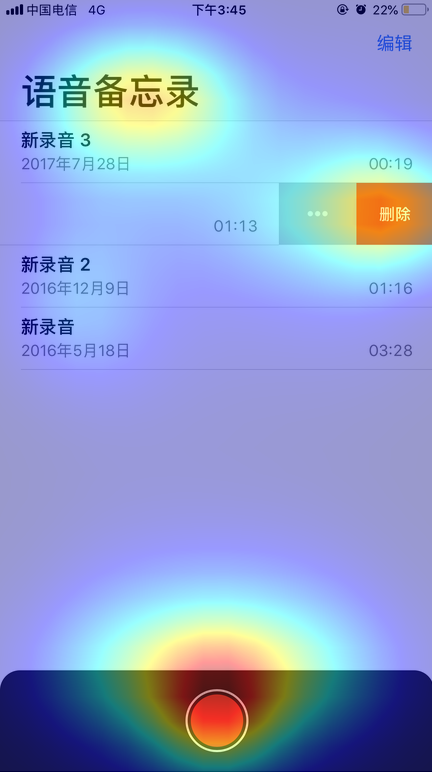} &
    \includegraphics[width=\w]{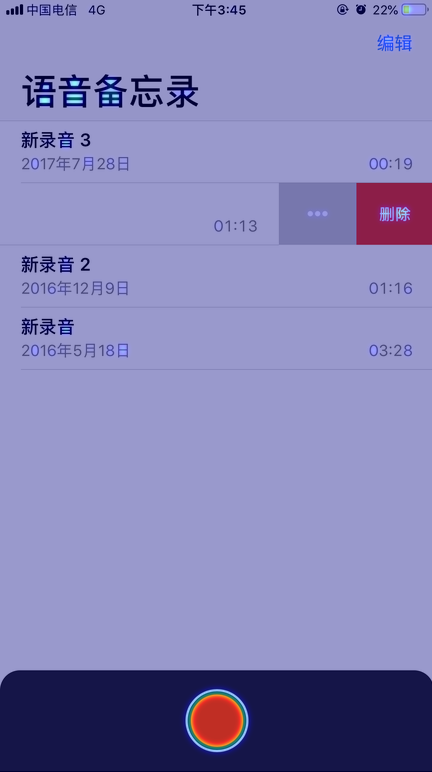} &
    \includegraphics[width=\w]{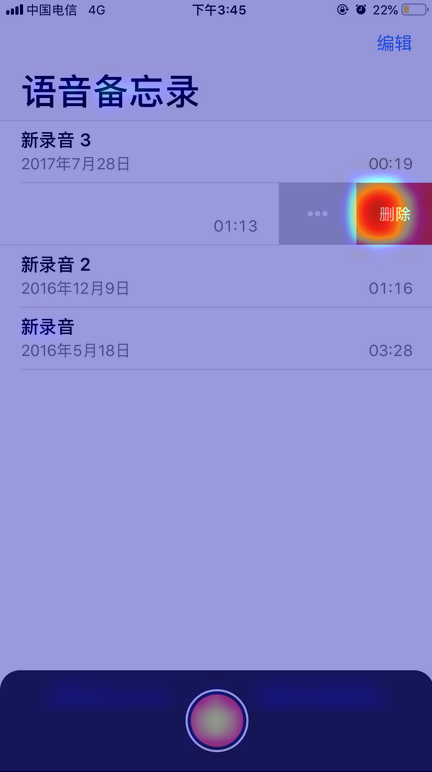} \\

    \tth{Source UI} & \tth{Ground-truth}
    & \tth{SAM-mobile} & \tth{SAM-S2015} & \tth{SAM-S2017} & \tth{ResNet-Sal} & \tth{GBVS} & \tth{BMS} & \tth{ITTI} \\

    \includegraphics[width=\w]{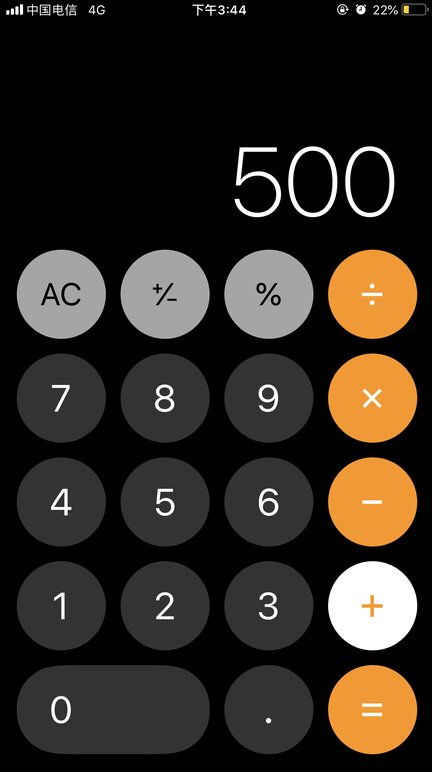} &
    \includegraphics[width=\w]{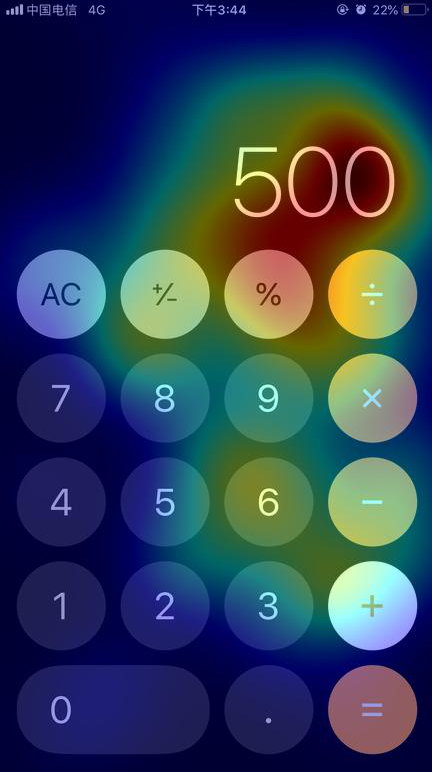} &
    \includegraphics[width=\w]{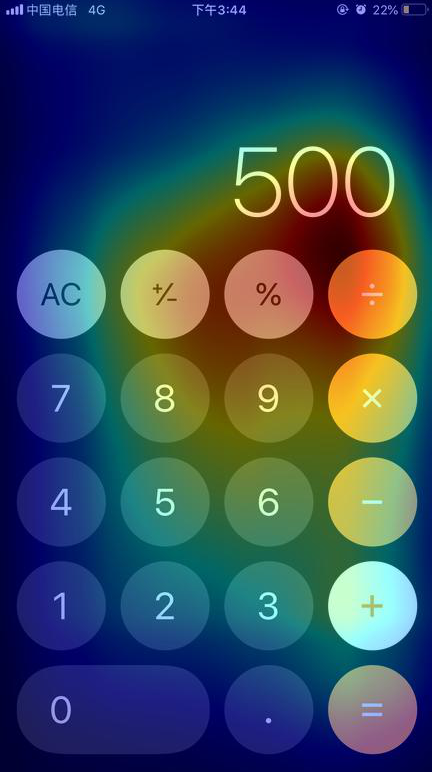} &
    \includegraphics[width=\w]{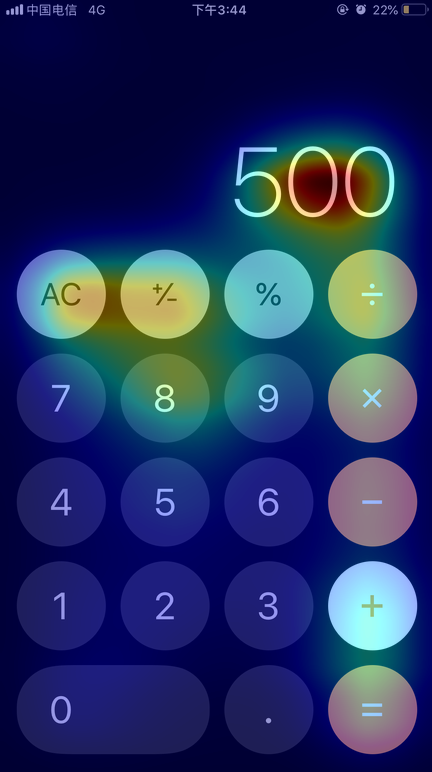} &
    \includegraphics[width=\w]{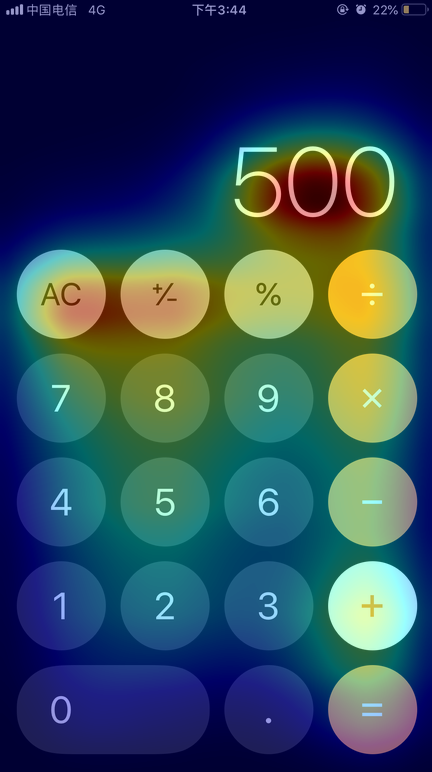} &
    \includegraphics[width=\w]{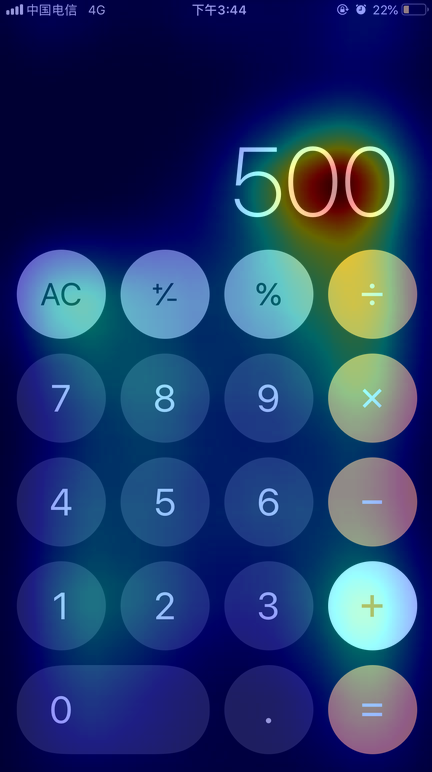} &
    \includegraphics[width=\w]{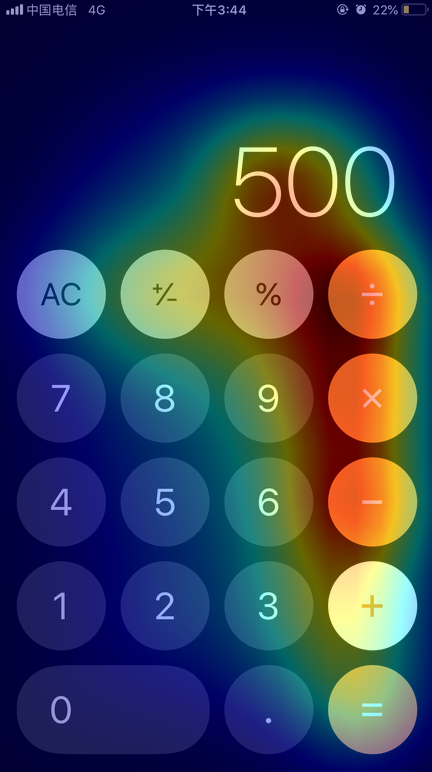} &
    \includegraphics[width=\w]{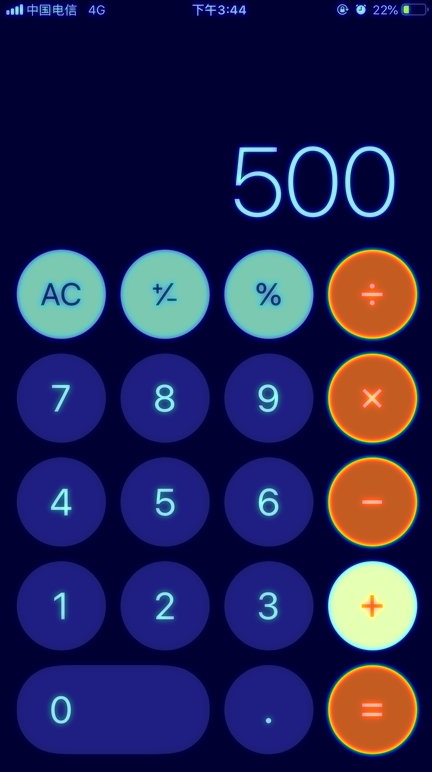} &
    \includegraphics[width=\w]{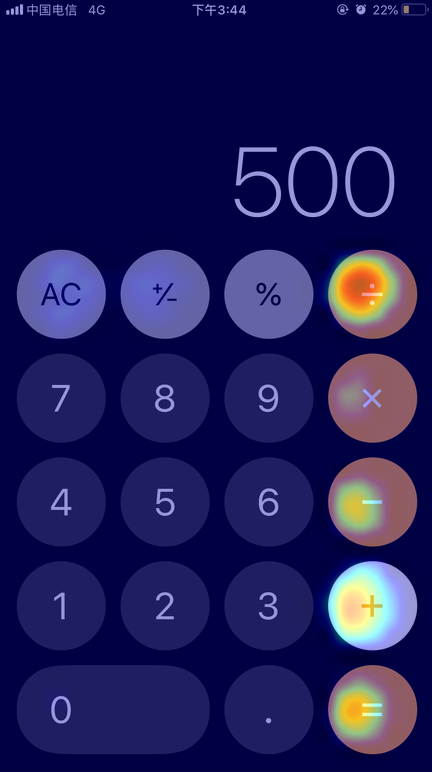} \\

    \tth{Source UI} & \tth{Ground-truth}
    & \tth{SAM-mobile} & \tth{SAM-S2015} & \tth{SAM-S2017} & \tth{ResNet-Sal} & \tth{GBVS} & \tth{BMS} & \tth{ITTI} \\

    \includegraphics[width=\w]{figs/saliency/hw7.png} &
    \includegraphics[width=\w]{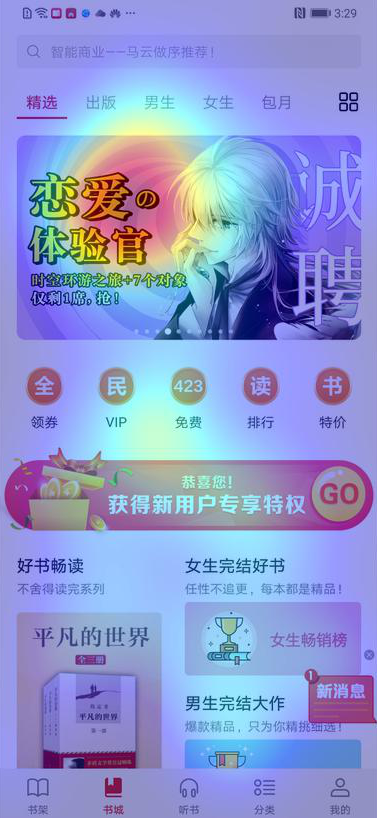} &
    \includegraphics[width=\w]{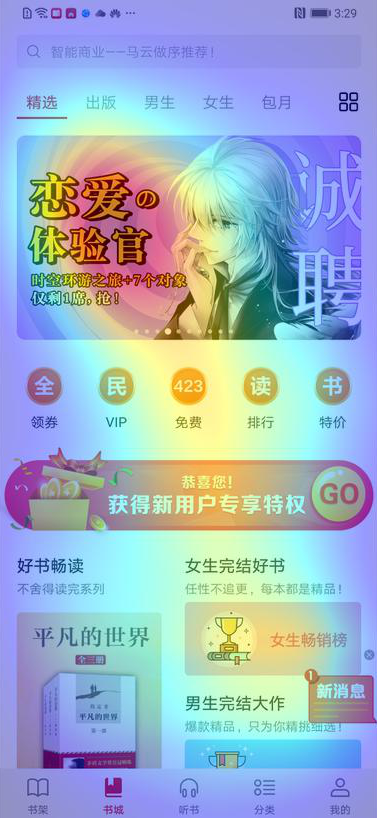} &
    \includegraphics[width=\w]{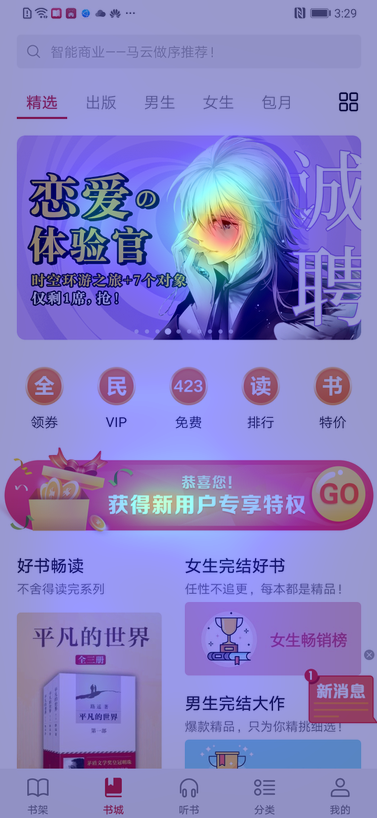} &
    \includegraphics[width=\w]{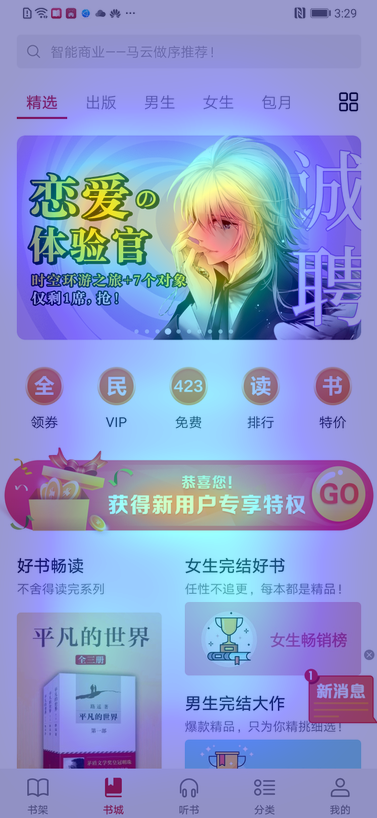} &
    \includegraphics[width=\w]{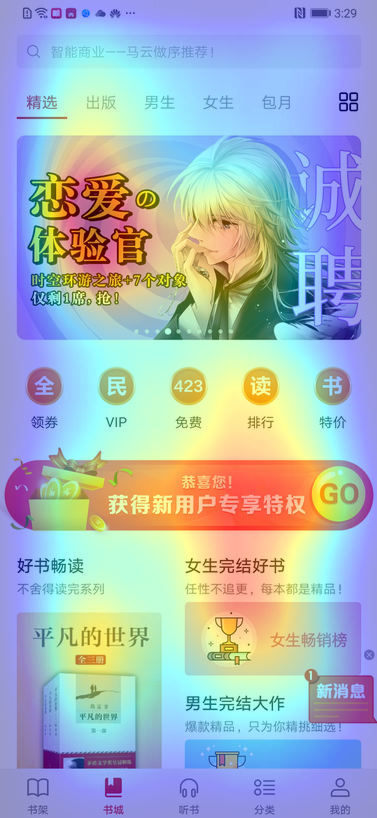} &
    \includegraphics[width=\w]{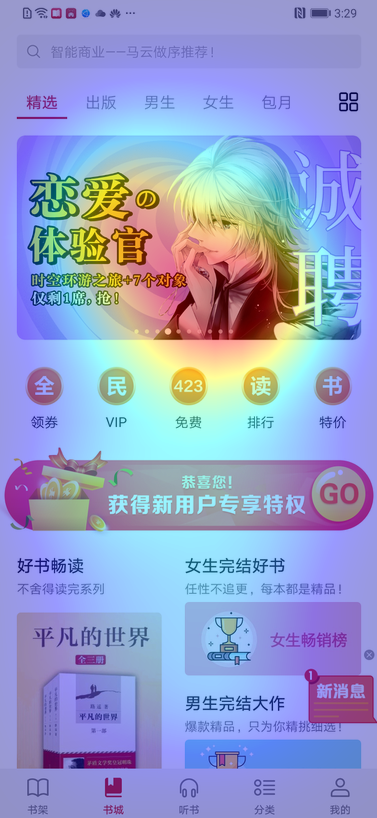} &
    \includegraphics[width=\w]{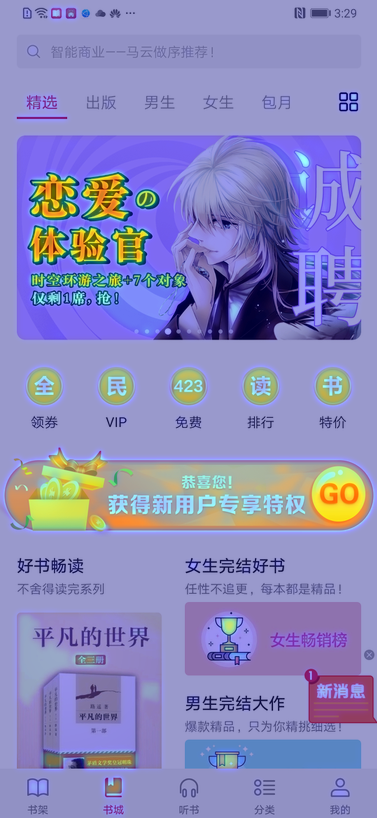} &
    \includegraphics[width=\w]{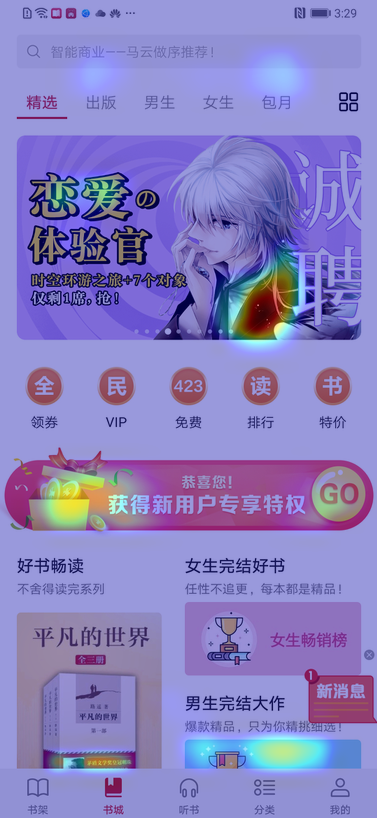} \\

    \tth{Source UI} & \tth{Ground-truth}
    & \tth{SAM-mobile} & \tth{SAM-S2015} & \tth{SAM-S2017} & \tth{ResNet-Sal} & \tth{GBVS} & \tth{BMS} & \tth{ITTI} \\

    \includegraphics[width=\w]{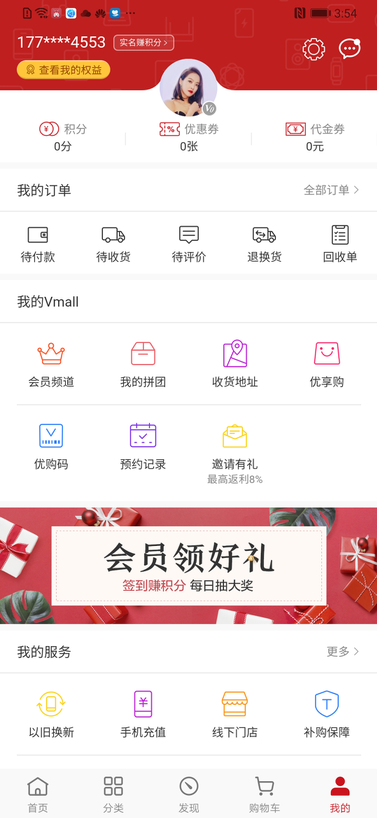} &
    \includegraphics[width=\w]{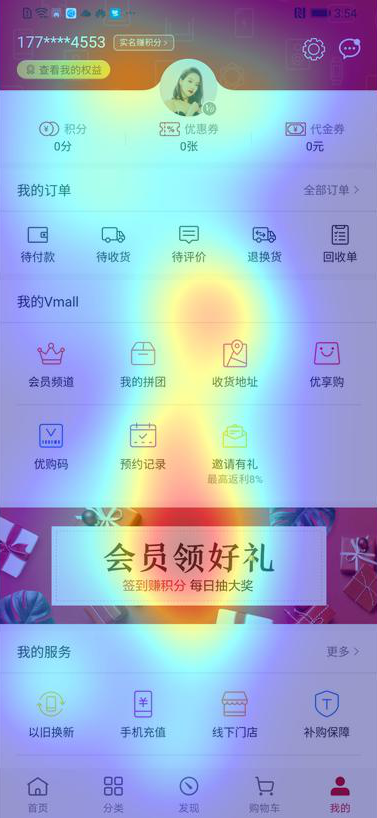} &
    \includegraphics[width=\w]{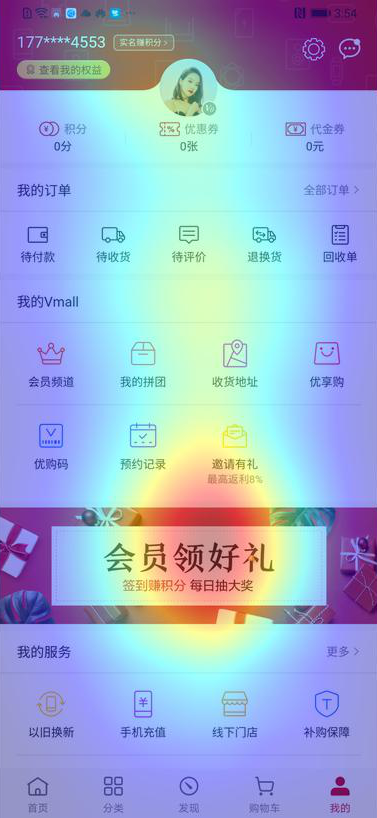} &
    \includegraphics[width=\w]{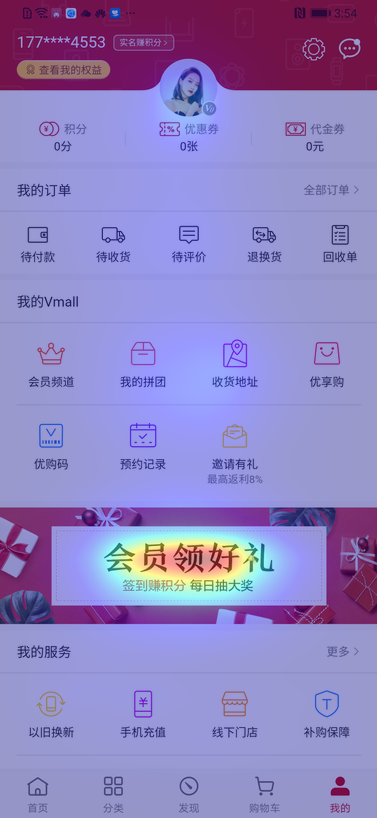} &
    \includegraphics[width=\w]{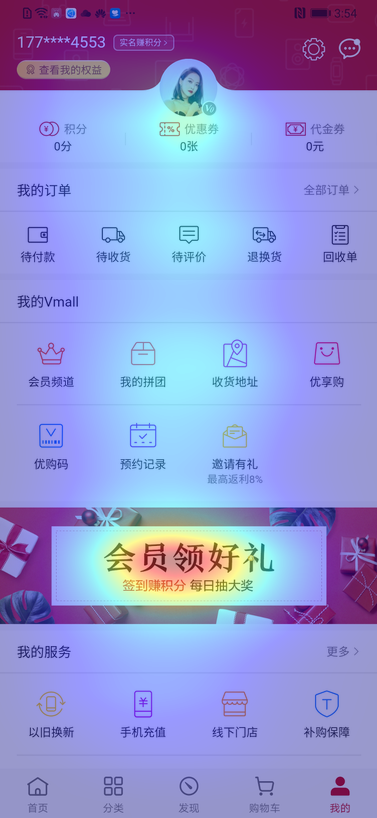} &
    \includegraphics[width=\w]{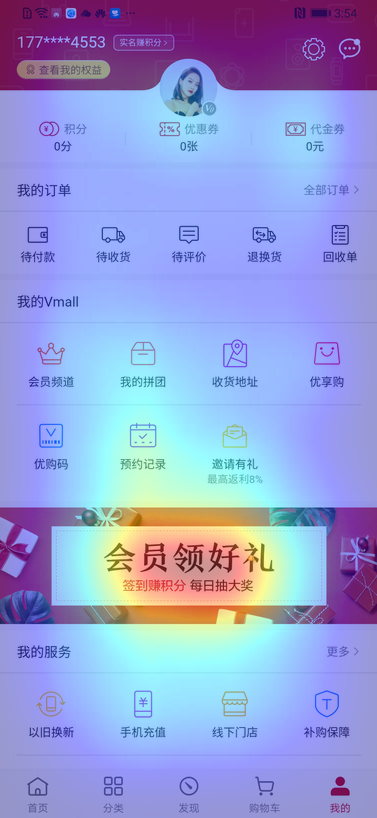} &
    \includegraphics[width=\w]{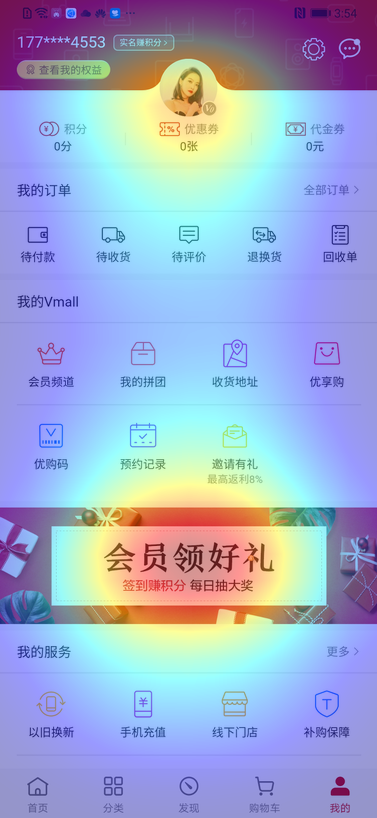} &
    \includegraphics[width=\w]{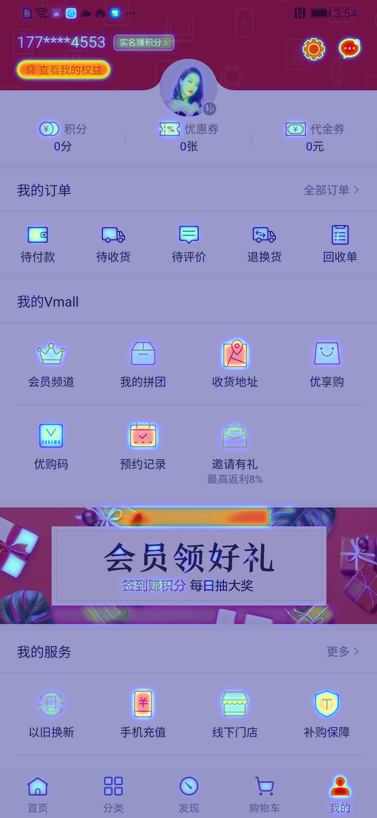} &
    \includegraphics[width=\w]{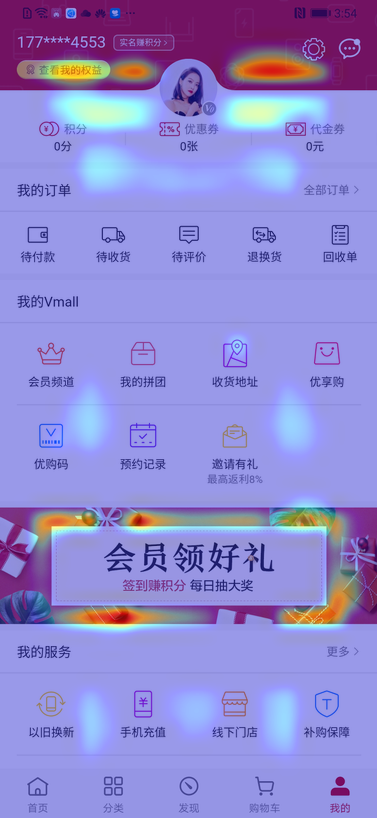} \\
  \end{tabular}

  \caption{Examples of saliency maps predicted for classic (ITTI, BMS, and GBVS) and modern (SAM and ResNet-Sal) models.}
  \label{fig:good-results}
  \Description{Several UI examples are shown with saliency heatmaps overlaid.}
\end{figure*}

\autoref{tbl:results} summarizes the modeling results,
and \autoref{fig:good-results} provides illustrative results from each saliency model.
We used the test for equality of proportions (with Yates' continuity correction)
to test the null hypothesis that all performance metrics provided by the models are the same.
The omnibus test was statistically significant for all metrics,
suggesting that some models performed significantly better than the others.
Therefore, we ran pairwise comparisons for equality of proportions as post-hoc test,
to better understand the differences between models.

\begin{table*}[!ht]
\centering
\begin{tabular}{ll *7c}
\toprule
\textbf{Metric} & \textbf{Type}
& \textbf{ITTI} & \textbf{BMS} & \textbf{GBVS}
& \textbf{ResNet-Sal}
& \textbf{SAM-S2015} & \textbf{SAM-S2017} & \textbf{SAM-mobile}  \\
\midrule
AUC & Location-based     & 0.223 & 0.249 & 0.666 & 0.692 & 0.650 & 0.666 & \bf 0.723 \\
NSS & \texttt{"}         & 0.126 & 0.138 & 0.591 & 0.704 & 0.537 & 0.655 & \bf 0.839 \\
SIM & Distribution-based & 0.558 & 0.206 & 0.709 & 0.734 & 0.562 & 0.664 & \bf 0.819 \\
 CC & \texttt{"}         & 0.082 & 0.131 & 0.580 & 0.657 & 0.477 & 0.621 & \bf 0.834 \\
\bottomrule
\end{tabular}
\caption{Results of our saliency models' performance.
The best result is highlighted in boldface type.}
\label{tbl:results}
\end{table*}

\textbf{AUC:} For area under the ROC curve,
both ITTI and BMS performed significantly worse than the other models:
$\chi^2_{(6, N=194)} = 357.177$, $p < .001$, $\phi = 1.357$.
No statistically significant differences were found between the other models.
We conclude that ResNet-Sal and all SAM variants show equally good predictive power.

\textbf{NSS:} For normalized scanpath saliency,
SAM-mobile performed significantly better than any of the other models:
$\chi^2_{(6, N=194)} = 357.177$, $p < .001$, $\phi = 1.357$.
No statistically significant differences were found between ITTI and BMS;
between GBVS and ResNet-Sal, SAM-S2015, or SAM-S2017;
between ResNet-Sal and SAM-S2017; or
between SAM-S2015 and SAM-S2017.
All other comparison results were found to be statistically significant.
We conclude that SAM-mobile yields the fixation predictions most closely aligned with ground-truth data.

\textbf{SIM:} For similarity,
SAM-mobile performed significantly better than all other models except GBVS and ResNet-Sal:
$\chi^2_{(6, N=194)} = 195.214$, $p < .001$, $\phi = 1.003$
No statistically significant differences were found between ITTI and SAM-S2015 or SAM-S2017,
between GBVS and ResNet-Sal or SAM-S2017,
or between ResNet-Sal and SAM-S2015 or SAM-S2017.
All other comparisons were found to be statistically significant.
We conclude that SAM-mobile, GBVS, and ResNet-Sal provide
equally good similarity between the predicted and ground-truth fixation maps.

\textbf{CC:} Finally, for the coefficient of correlation,
SAM-mobile performed significantly better than any of the other models:
$\chi^2_{(6, N=194)} = 362.549$, $p < .001$, $\phi = 1.367$.
No statistically significant differences were found between ITTI and BMS,
between GBVS and ResNet-Sal, SAM-S2015, and SAM-S2017,
and between ResNet-Sal and SAM-S2017.
All other comparisons showed statistical significance.
We conclude that SAM-mobile is the model that correlates the best with human visual saliency.

\subsubsection{Summary}

\begin{figure}[!ht]
  \centering
  \setlength{\tabcolsep}{2pt}
  \def\w{0.11\linewidth}
  \begin{tabular}{*5c}
    \tth{Source UI} & \tth{Ground truth} & \tth{SAM-mobile} & \tth{SAM-S2015} & \tth{SAM-S2017} \\

    \includegraphics[width=\w]{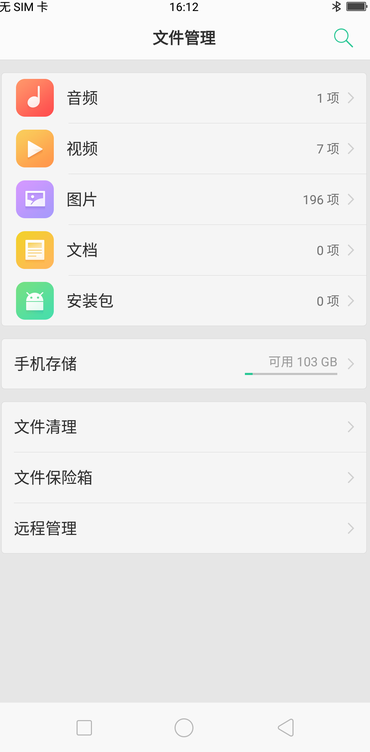} &
    \includegraphics[width=\w]{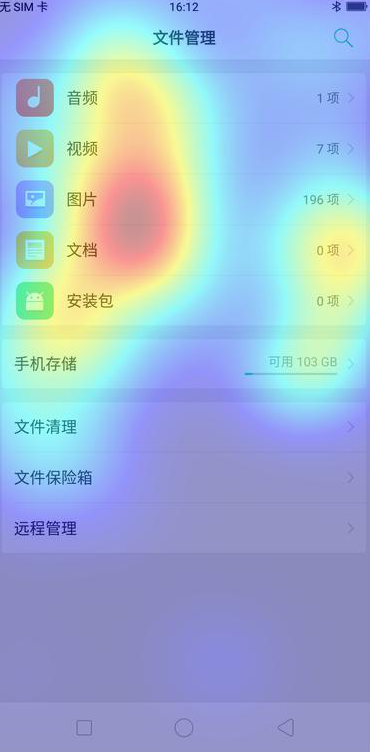} &
    \includegraphics[width=\w]{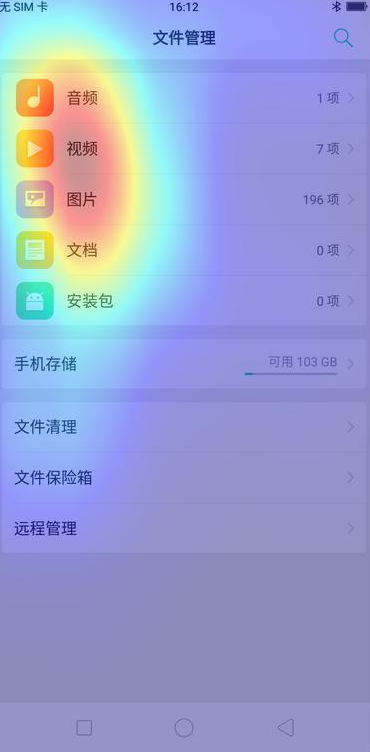} &
    \includegraphics[width=\w]{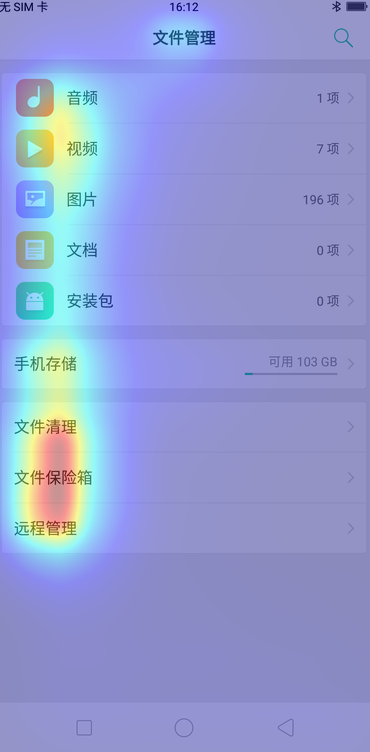} &
    \includegraphics[width=\w]{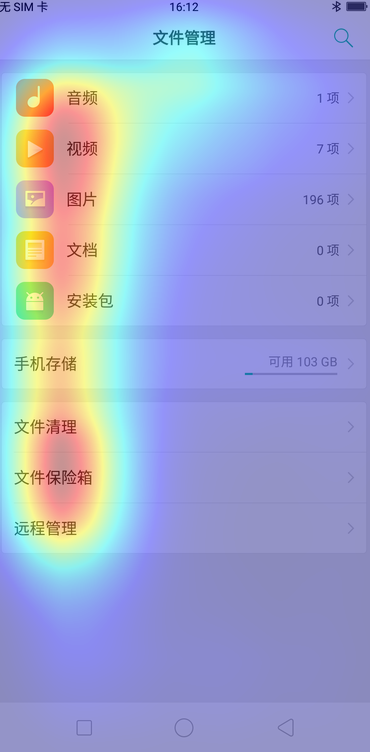} \\

    \includegraphics[width=\w]{figs/saliency/hw3.png} &
    \includegraphics[width=\w]{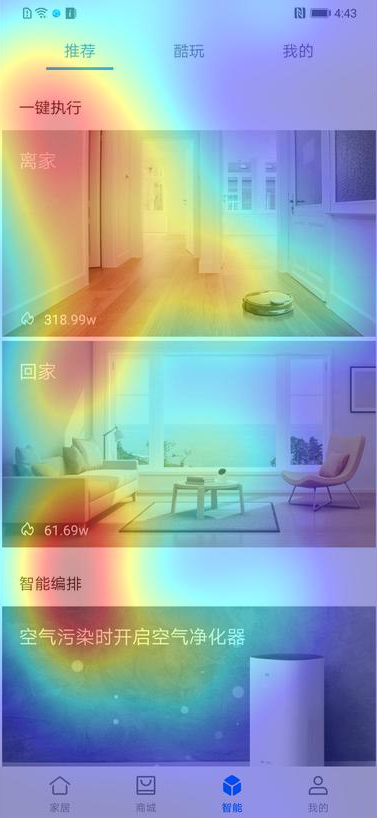} &
    \includegraphics[width=\w]{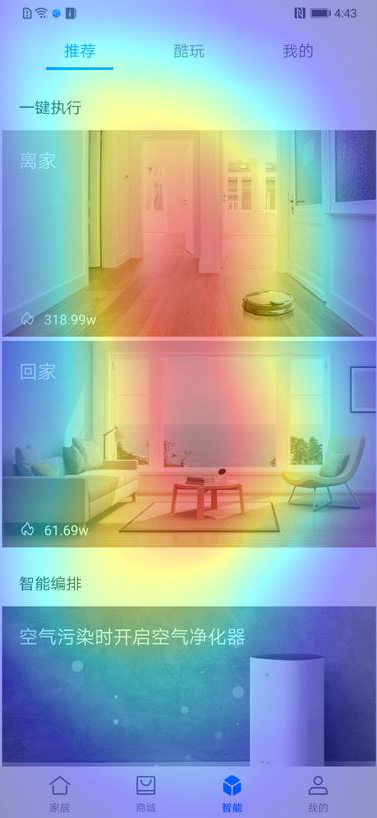} &
    \includegraphics[width=\w]{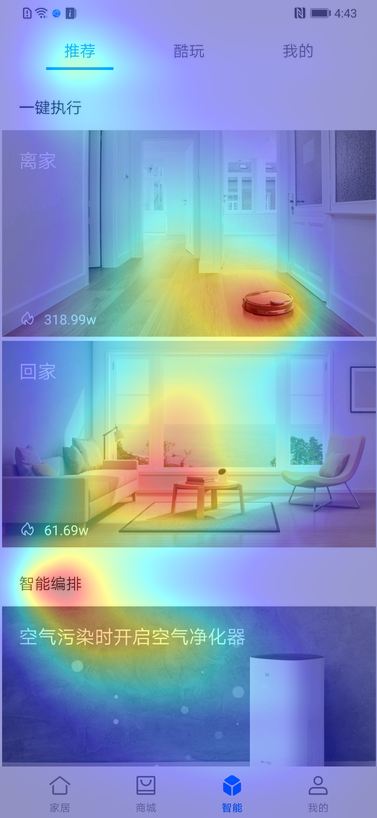} &
    \includegraphics[width=\w]{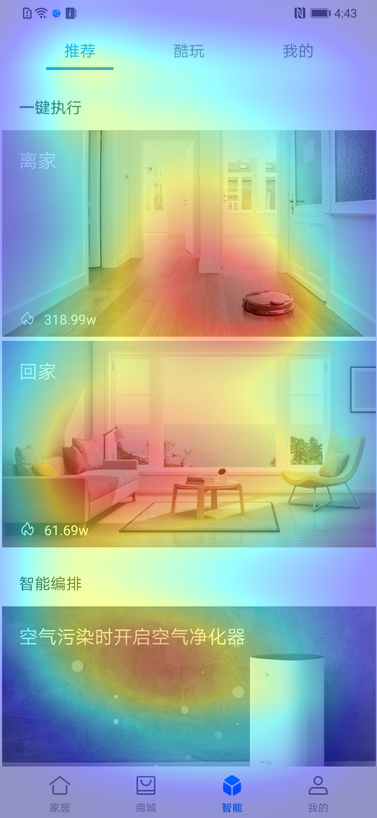} \\
  \end{tabular}
  \caption{Interfaces with poor prediction -- examples from SAM-related models.}
  \label{fig:bad-results}
  \Description{Several UI examples are shown with saliency heatmaps overlaid.}
\end{figure}

Stimulus-driven models such as ITTI and BMS performed worse in general,
with data-driven models generally achieving a higher fit.
SAM-mobile emerges as the overall winner with these evaluation metrics.
We discuss our explanation for this in the next section of the paper.
\autoref{fig:bad-results} illustrates some cases of the models failing to predict visual saliency accurately.
While the general accuracy is high, this gallery suggests that there is still some room for improvement.

\section{Discussion}
\label{sec:discussion}
This paper has presented an in-depth study of visual saliency in mobile interfaces.
Sampled from numerous Android and iOS applications,
the interfaces varied in style, content, and structure.
The material we collected is the first representative and comprehensive dataset for this topic.
We assessed known empirical effects and tested several classical and state-of-the-art saliency models against the fixation data collected.

In reporting the first empirical findings from this dataset,
we have focused on testing some known saliency biases.
Our key findings from comparing mobile UIs to natural scenes are the following:
\begin{enumerate}
\item Strong location bias toward the top-left quadrant of the screen
\item Strong text bias
\item Strong image bias
\item Strong face bias
\item No effect of element size
\item Neither center nor horizontal bias
\item No color bias
\end{enumerate}

Location had the strongest effect on saliency in our data,
with the top-left quadrant (Q2) dominating the fixation data.
This has not been reported before for mobile devices,
although the effect is predictable.
F-shaped and triangular gaze patterns have been reported for webpages~\cite{Duggan11}.
We believe the most plausible reason is statistical learning of visual-feature distribution.
In other words, designers prioritize the upper left for information that helps orient the user.
Logos, headers, and instructions are most likely to reside in that region.
Moreover, labels, text, and headings are often read from top to bottom, left to right.
With repeated exposure to designs of this type,
users learn that the top-left corner makes a wise fixation choice.

It was somewhat surprising to find a lack of bias based on color or size,
since these are elementary stimulus features considered in bottom-up saliency models.
There is an explanation, though.
If the distributions of these features are uninformative,
the cue cannot be exploited for saliency,
and users learn to avoid them.
It could be that the size of an element, similarly, is not a useful cue for identifying which elements are informative.
Consider the wide distribution we found for colors in the UIs.
The only prominent color is gray, which is prevalent in backgrounds.

In contrast, it was less surprising that text and images attract attention,
since they have a role in a user's orientation.
The text bias attests to the visual system's ability to utilize visual cues that hint at the semantics of an object,
even when as inconspicuous as text elements are.
Images, on the other hand, are typically placed in a UI with the goal of attracting attention,
with the colors, objects etc. chosen to stand out.

Taken together, these findings suggest that users exhibit viewing strategies
tuned to their expectations of the feature distributions of mobile UIs.
Indeed, UIs are accessed by users for specific purposes, with particular interactions and a goal in mind.
We argue that visual statistical learning may help explain the thrust of these results,
which could further help explain the modeling output.

A more general finding is that classic models performed worse than the data-driven deep learning models.
This might reflect the fact that the classic models are based on stimulus-driven computations,
whereas the data-driven models were trained such that they had access to data on users' viewing strategies.
Across all metrics we examined, the SAM-mobile deep learning model achieved the best performance.
Other models came close but never surpassed it.
The superiority of the SAM-mobile model is explained by
(1) the attentive saliency decoding network and
(2) the use of transfer learning for the mobile dataset.
The latter case is particularly important,
since classic models of saliency were mainly developed for natural images
and the vast majority of DL models of visual saliency are trained and evaluated with natural images.

If users' viewing strategies are indeed tuned to particular distributions of features,
a model trained specifically for the relevant type of data is bound to have a better fit.
Why, then, was ResNet-Sal, which was trained on the same data and fine-tuned to the mobile domain,
performed worse than SAM-mobile?
We attribute the difference to SAM-mobile's attentive saliency decoding network,
which may be better at learning strategies of overt attention from data.
Related to this, we noticed that many fixations happened on backgrounds,
which can be attributed to covert attention cases
since our eye-tracker was re-calibrated for each participant.
In covert attention cases, the user is attending to an object
without precisely placing the eye gaze on that object.

Overall, our modeling results will benefit practitioners who routinely employ saliency models.
In light of the results, we conclude that all deep-learning-based models provide reasonable accuracy off-the-shelf.
However, as our study shows, accuracy can be further improved by training from a mobile-specific dataset.
While this observation has been highlighted by others~\cite{Borji19, Bylinskii17},
until now it was not possible to analyze saliency in mobile UIs
because of the lack of a public dataset.
Our results together fuel a recommendation that practitioners should calibrate their models to this domain.

\subsection{Limitations and Future Work}

More work is needed for refining our understanding of saliency in the mobile domain.
For example, our study should be replicated in task environments with users who are truly mobile.
Paradigmatically, visual saliency has been studied under free-viewing (bottom-up) conditions as opposed to specific goals (top-down).
While quantifying bottom-up saliency, user's familiarity with various mobile platforms plays a lesser important role,
however we suspect that it may play a key role while quantifying top-down saliency.
Bottom-up saliency is frequently considered by designers and researchers (see e.g. \cite{Bylinskii17}),
so this paper contributes important information on the unique aspects of mobile UIs.
Top-down saliency is affected by expectations, location memory, search strategies, among other fenomena,
which is a complex topic worthy of future work.

We note also that, according to previous work~\cite{Chua05, Kelly10},
Eastern people often look at and perceive objects differently from Westerners.
Chua et al.~\cite{Chua05} found that the latter attend more to focal objects in natural scenes,
while East Asians pay more attention to contextual information,
and Kelly et al.~\cite{Kelly10} reported that East Asian subjects displayed central fixations across all categories of visual stimuli.
To investigate these differences between holistic and analytic vision, further research is warranted.

In addition, our saliency models predict a static global distribution of visual attention over the image,
ignoring the active role of attention in planning the next fixation point.
We believe that models combining goal-driven, learned, and bottom-up features
are required for enhancing predictions of the sequence of fixation points.
Future work should also consider the role of individuals' differences
as well as visual search strategies, which are mostly guided by top-down saliency principles~\cite{Jokinen20}.

\section{Conclusion}
\label{sec:conclusion}
Our work has shown that the common understanding of visual saliency,
especially with regard to some prominent effects identified in studies of natural scenes,
does not transfer trivially to mobile UIs.
In particular, we learned that saliency is dominated by location and semantic biases,
which we believe to be due to their informativeness as conspicuous cues.
In contrast, it is dominated less by some of the core bottom-up features, such as size and color,
perhaps reflecting that those feature distributions tend to be less informative in this domain.
For designers, our results mean that it is better to follow the style and conventions of the domain
in, for instance, the choice of where to place important elements.
Even conspicuous cues, such as use of red or large size, may go unnoticed,
given that mobile UIs are rich in features of this sort.
That said, additional analyses could exploit our data, for better understanding of this phenomena.
For example, one could look at the interaction effect between visual features and element types.
Do images grab attention more because they are more colorful?
What kinds of color distributions do we see in the four quadrants?
Our observations highlight a need for more research into what, if anything,
makes mobile UIs special.

\section{The `\textsc{Mobile UI Saliency}' Dataset}
\label{sec:dataset}
The outcome of our study is a rich annotated dataset characterizing the attraction of visual attention in mobile interfaces.
We are publicly releasing the dataset, which includes:
\begin{itemize}
\item Screenshots of 193 mobile UIs sampled from present-day application markets.
\item Annotations (bounding boxes) of all UI elements in the screenshots.
\item Eye-tracking data, including fixation points, timestamps, and aggregated indices like heatmaps.
\item The full list of mobile apps and vendors.
\end{itemize}

\begin{acks}
We thank Marko Repo for his help with UI element annotation
and the anonymous referees for their feedback.
We also thank the computational resources provided by the Aalto Science-IT project.
This work has been supported by the Academy of Finland (grant no. 318559),
the Huawei UCD Center,
and the European Research Council (ERC) under the European Union's Horizon 2020 research
and innovation programme (grant agreement no. 637991).
\end{acks}


\begin{thebibliography}{68}

\ifx \showCODEN    \undefined \def \showCODEN     #1{\unskip}     \fi
\ifx \showDOI      \undefined \def \showDOI       #1{#1}\fi
\ifx \showISBNx    \undefined \def \showISBNx     #1{\unskip}     \fi
\ifx \showISBNxiii \undefined \def \showISBNxiii  #1{\unskip}     \fi
\ifx \showISSN     \undefined \def \showISSN      #1{\unskip}     \fi
\ifx \showLCCN     \undefined \def \showLCCN      #1{\unskip}     \fi
\ifx \shownote     \undefined \def \shownote      #1{#1}          \fi
\ifx \showarticletitle \undefined \def \showarticletitle #1{#1}   \fi
\ifx \showURL      \undefined \def \showURL       {\relax}        \fi
\providecommand\bibfield[2]{#2}
\providecommand\bibinfo[2]{#2}
\providecommand\natexlab[1]{#1}
\providecommand\showeprint[2][]{arXiv:#2}

\bibitem[\protect\citeauthoryear{Apple}{Apple}{2019}]
        {Apple:guidelines}
\bibfield{author}{\bibinfo{person}{Apple}.} \bibinfo{year}{2019}\natexlab{}.
\newblock \bibinfo{title}{Human Interface Guidelines: iOS}.
\newblock \bibinfo{howpublished}{Available:
  \url{https://developer.apple.com/design/human-interface-guidelines/ios/overview/themes/}
  (last accessed January 2020)}.
\newblock

\bibitem[\protect\citeauthoryear{Bezryadin, Bourov, and Ilinih}{Bezryadin
  et~al\mbox{.}}{2007}]
        {Bezryadin07}
\bibfield{author}{\bibinfo{person}{S. Bezryadin}, \bibinfo{person}{P. Bourov},
  {and} \bibinfo{person}{D. Ilinih}.} \bibinfo{year}{2007}\natexlab{}.
\newblock \showarticletitle{Brightness Calculation in Digital Image
  Processing}. In \bibinfo{booktitle}{\emph{Proc. TDPF Symposium}}.
\newblock

\bibitem[\protect\citeauthoryear{Borji}{Borji}{2019}]
        {Borji19}
\bibfield{author}{\bibinfo{person}{A. Borji}.} \bibinfo{year}{2019}\natexlab{}.
\newblock \showarticletitle{Saliency Prediction in the Deep Learning Era:
  Successes, Limitations, and Future Challenges}. In
  \bibinfo{booktitle}{\emph{CoRR abs/1810.03716 (arXiv preprint)}}.
\newblock

\bibitem[\protect\citeauthoryear{Borji and Itti}{Borji and Itti}{2013}]
        {Borji13}
\bibfield{author}{\bibinfo{person}{A. Borji} {and} \bibinfo{person}{L. Itti}.}
  \bibinfo{year}{2013}\natexlab{}.
\newblock \showarticletitle{State-of-the-art in visual attention modeling}.
\newblock \bibinfo{journal}{\emph{IEEE Trans. Pattern Anal. Mach. Intell.}}
  \bibinfo{volume}{35}, \bibinfo{number}{1} (\bibinfo{year}{2013}).
\newblock

\bibitem[\protect\citeauthoryear{Borji, Tavakoli, Sihite, and Itti}{Borji
  et~al\mbox{.}}{2013}]
        {Borji_2013_ICCV}
\bibfield{author}{\bibinfo{person}{A. Borji}, \bibinfo{person}{H.~R. Tavakoli},
  \bibinfo{person}{D.~N. Sihite}, {and} \bibinfo{person}{L. Itti}.}
  \bibinfo{year}{2013}\natexlab{}.
\newblock \showarticletitle{Analysis of Scores, Datasets, and Models in Visual
  Saliency Prediction}. In \bibinfo{booktitle}{\emph{Proc. ICCV}}.
\newblock

\bibitem[\protect\citeauthoryear{Bylinskii, Judd, Oliva, Torralba, and
  Durand}{Bylinskii et~al\mbox{.}}{2019}]
        {Bylinskii19}
\bibfield{author}{\bibinfo{person}{Z. Bylinskii}, \bibinfo{person}{T. Judd},
  \bibinfo{person}{A. Oliva}, \bibinfo{person}{A. Torralba}, {and}
  \bibinfo{person}{F. Durand}.} \bibinfo{year}{2019}\natexlab{}.
\newblock \showarticletitle{What Do Different Evaluation Metrics Tell Us About
  Saliency Models?}
\newblock \bibinfo{journal}{\emph{IEEE Trans. Pattern Anal. Mach. Intell.}}
  \bibinfo{volume}{41}, \bibinfo{number}{3} (\bibinfo{year}{2019}).
\newblock

\bibitem[\protect\citeauthoryear{Bylinskii, Kim, O'Donovan, Alsheikh, Madan,
  Pfister, Durand, Russell, and Hertzmann}{Bylinskii et~al\mbox{.}}{2017}]
        {Bylinskii17}
\bibfield{author}{\bibinfo{person}{Z. Bylinskii}, \bibinfo{person}{N.~W. Kim},
  \bibinfo{person}{P. O'Donovan}, \bibinfo{person}{S. Alsheikh},
  \bibinfo{person}{S. Madan}, \bibinfo{person}{H. Pfister}, \bibinfo{person}{F.
  Durand}, \bibinfo{person}{B. Russell}, {and} \bibinfo{person}{A. Hertzmann}.}
  \bibinfo{year}{2017}\natexlab{}.
\newblock \showarticletitle{Learning Visual Importance for Graphic Designs and
  Data Visualizations}. In \bibinfo{booktitle}{\emph{Proc. UIST}}.
\newblock

\bibitem[\protect\citeauthoryear{Cerf, Frady, and Koch}{Cerf
  et~al\mbox{.}}{2009}]
        {Cerf2009}
\bibfield{author}{\bibinfo{person}{M. Cerf}, \bibinfo{person}{E.~P. Frady},
  {and} \bibinfo{person}{C. Koch}.} \bibinfo{year}{2009}\natexlab{}.
\newblock \showarticletitle{Faces and text attract gaze independent of the
  task: Experimental data and computer model}.
\newblock \bibinfo{journal}{\emph{J. Vis.}} \bibinfo{volume}{9},
  \bibinfo{number}{12} (\bibinfo{year}{2009}).
\newblock

\bibitem[\protect\citeauthoryear{Chua, Boland, and Nisbett}{Chua
  et~al\mbox{.}}{2005}]
        {Chua05}
\bibfield{author}{\bibinfo{person}{H.~F. Chua}, \bibinfo{person}{J.~E. Boland},
  {and} \bibinfo{person}{R.~E. Nisbett}.} \bibinfo{year}{2005}\natexlab{}.
\newblock \showarticletitle{Cultural variation in eye movements during scene
  perception}.
\newblock \bibinfo{journal}{\emph{PNAS}} \bibinfo{volume}{102},
  \bibinfo{number}{35} (\bibinfo{year}{2005}).
\newblock

\bibitem[\protect\citeauthoryear{Cooke}{Cooke}{2006}]
        {Cooke06}
\bibfield{author}{\bibinfo{person}{L. Cooke}.} \bibinfo{year}{2006}\natexlab{}.
\newblock \showarticletitle{Is the mouse a poor man's eye tracker?}. In
  \bibinfo{booktitle}{\emph{Proc. STC}}.
\newblock

\bibitem[\protect\citeauthoryear{{Cornia}, {Baraldi}, {Serra}, and
  {Cucchiara}}{{Cornia} et~al\mbox{.}}{2018}]
        {Cornia18}
\bibfield{author}{\bibinfo{person}{M. {Cornia}}, \bibinfo{person}{L.
  {Baraldi}}, \bibinfo{person}{G. {Serra}}, {and} \bibinfo{person}{R.
  {Cucchiara}}.} \bibinfo{year}{2018}\natexlab{}.
\newblock \showarticletitle{Predicting Human Eye Fixations via an LSTM-Based
  Saliency Attentive Model}.
\newblock \bibinfo{journal}{\emph{IEEE Trans. Image Process.}}
  \bibinfo{volume}{27}, \bibinfo{number}{10} (\bibinfo{year}{2018}).
\newblock

\bibitem[\protect\citeauthoryear{Duggan and Payne}{Duggan and Payne}{2011}]
        {Duggan11}
\bibfield{author}{\bibinfo{person}{G.~B. Duggan} {and} \bibinfo{person}{S.~J.
  Payne}.} \bibinfo{year}{2011}\natexlab{}.
\newblock \showarticletitle{Skim Reading by Satisficing: Evidence from Eye
  Tracking}. In \bibinfo{booktitle}{\emph{Proc. CHI}}.
\newblock

\bibitem[\protect\citeauthoryear{Etchebehere and Fedorovskaya}{Etchebehere and
  Fedorovskaya}{2017}]
        {Etchebehere17}
\bibfield{author}{\bibinfo{person}{S. Etchebehere} {and} \bibinfo{person}{E.
  Fedorovskaya}.} \bibinfo{year}{2017}\natexlab{}.
\newblock \showarticletitle{On the Role of Color in Visual Saliency}.
\newblock \bibinfo{journal}{\emph{Intl. Symp. Electronic Imaging}}
  \bibinfo{volume}{6} (\bibinfo{year}{2017}).
\newblock

\bibitem[\protect\citeauthoryear{Frintrop, Rome, and Christensen}{Frintrop
  et~al\mbox{.}}{2010}]
        {Frintrop10}
\bibfield{author}{\bibinfo{person}{S. Frintrop}, \bibinfo{person}{E. Rome},
  {and} \bibinfo{person}{H.~I. Christensen}.} \bibinfo{year}{2010}\natexlab{}.
\newblock \showarticletitle{Computational visual attention systems and their
  cognitive foundations: A survey}.
\newblock \bibinfo{journal}{\emph{ACM Trans. Appl. Percept.}}
  \bibinfo{volume}{7}, \bibinfo{number}{1} (\bibinfo{year}{2010}).
\newblock

\bibitem[\protect\citeauthoryear{Goodfellow, Bengio, and Courville}{Goodfellow
  et~al\mbox{.}}{2016}]
        {Goodfellow16}
\bibfield{author}{\bibinfo{person}{I. Goodfellow}, \bibinfo{person}{Y. Bengio},
  {and} \bibinfo{person}{A. Courville}.} \bibinfo{year}{2016}\natexlab{}.
\newblock \bibinfo{booktitle}{\emph{Deep Learning}}.
\newblock

\bibitem[\protect\citeauthoryear{Google}{Google}{2019}]
        {Google:guidelines}
\bibfield{author}{\bibinfo{person}{Google}.} \bibinfo{year}{2019}\natexlab{}.
\newblock \bibinfo{title}{Google Material Guidelines}.
\newblock \bibinfo{howpublished}{Available: \url{https://material.io/design/}
  (last accessed January 2020)}.
\newblock

\bibitem[\protect\citeauthoryear{Gupta, Gupta, Jayagopal, Pal, and Sinha}{Gupta
  et~al\mbox{.}}{2018}]
        {Gupta18}
\bibfield{author}{\bibinfo{person}{P. Gupta}, \bibinfo{person}{S. Gupta},
  \bibinfo{person}{A. Jayagopal}, \bibinfo{person}{S. Pal}, {and}
  \bibinfo{person}{R. Sinha}.} \bibinfo{year}{2018}\natexlab{}.
\newblock \showarticletitle{Saliency Prediction for Mobile User Interfaces}. In
  \bibinfo{booktitle}{\emph{Proc. WACV Workshop}}.
\newblock

\bibitem[\protect\citeauthoryear{Hamel, Guyader, Pellerin, and Houzet}{Hamel
  et~al\mbox{.}}{2014}]
        {Hamel14}
\bibfield{author}{\bibinfo{person}{S. Hamel}, \bibinfo{person}{N. Guyader},
  \bibinfo{person}{D. Pellerin}, {and} \bibinfo{person}{D. Houzet}.}
  \bibinfo{year}{2014}\natexlab{}.
\newblock \showarticletitle{Contribution of Color Information in Visual
  Saliency Model for Videos}. In \bibinfo{booktitle}{\emph{Proc. ICISP}}.
\newblock

\bibitem[\protect\citeauthoryear{Harel, Koch, and Perona}{Harel
  et~al\mbox{.}}{2007}]
        {Harel07}
\bibfield{author}{\bibinfo{person}{J. Harel}, \bibinfo{person}{C. Koch}, {and}
  \bibinfo{person}{P. Perona}.} \bibinfo{year}{2007}\natexlab{}.
\newblock \showarticletitle{Graph-Based Visual Saliency}. In
  \bibinfo{booktitle}{\emph{Proc. NIPS}}.
\newblock

\bibitem[\protect\citeauthoryear{He, Zhang, Ren, and Sun}{He
  et~al\mbox{.}}{2016}]
        {He16}
\bibfield{author}{\bibinfo{person}{K. He}, \bibinfo{person}{X. Zhang},
  \bibinfo{person}{S. Ren}, {and} \bibinfo{person}{J. Sun}.}
  \bibinfo{year}{2016}\natexlab{}.
\newblock \showarticletitle{Deep Residual Learning for Image Recognition}. In
  \bibinfo{booktitle}{\emph{Proc. CVPR}}.
\newblock

\bibitem[\protect\citeauthoryear{He, Tavakoli, Borji, Mi, and Pugeault}{He
  et~al\mbox{.}}{2019}]
        {He_2019_CVPR}
\bibfield{author}{\bibinfo{person}{S. He}, \bibinfo{person}{H.~R. Tavakoli},
  \bibinfo{person}{A. Borji}, \bibinfo{person}{Y. Mi}, {and}
  \bibinfo{person}{N. Pugeault}.} \bibinfo{year}{2019}\natexlab{}.
\newblock \showarticletitle{Understanding and Visualizing Deep Visual Saliency
  Models}. In \bibinfo{booktitle}{\emph{Proc. CVPR}}.
\newblock

\bibitem[\protect\citeauthoryear{Henderson}{Henderson}{1993}]
        {Henderson}
\bibfield{author}{\bibinfo{person}{J.~M. Henderson}.}
  \bibinfo{year}{1993}\natexlab{}.
\newblock \showarticletitle{Eye movement control during visual object
  processing: effects of initial fixation position and semantic constraint}.
\newblock \bibinfo{journal}{\emph{Can. J. Exp. Psychol.}} \bibinfo{volume}{47},
  \bibinfo{number}{1} (\bibinfo{year}{1993}).
\newblock

\bibitem[\protect\citeauthoryear{Humphrey and Underwood}{Humphrey and
  Underwood}{2012}]
        {Humphrey2012}
\bibfield{author}{\bibinfo{person}{K. Humphrey} {and} \bibinfo{person}{G.
  Underwood}.} \bibinfo{year}{2012}\natexlab{}.
\newblock \showarticletitle{The potency of people in pictures: Evidence from
  sequences of eye fixations}.
\newblock \bibinfo{journal}{\emph{J. Vis.}} \bibinfo{volume}{12},
  \bibinfo{number}{6} (\bibinfo{year}{2012}).
\newblock

\bibitem[\protect\citeauthoryear{Itti}{Itti}{2007}]
        {Itti:2007}
\bibfield{author}{\bibinfo{person}{L. Itti}.} \bibinfo{year}{2007}\natexlab{}.
\newblock \showarticletitle{{V}isual salience}.
\newblock \bibinfo{journal}{\emph{Scholarpedia}} \bibinfo{volume}{2},
  \bibinfo{number}{9} (\bibinfo{year}{2007}).
\newblock

\bibitem[\protect\citeauthoryear{Itti, Koch, and Niebur}{Itti
  et~al\mbox{.}}{1998}]
        {Itti98}
\bibfield{author}{\bibinfo{person}{L. Itti}, \bibinfo{person}{C. Koch}, {and}
  \bibinfo{person}{E. Niebur}.} \bibinfo{year}{1998}\natexlab{}.
\newblock \showarticletitle{A model of saliency based visual attention for
  rapid scene analysis}.
\newblock \bibinfo{journal}{\emph{IEEE Trans. Pattern Anal. Mach. Intell.}}
  \bibinfo{volume}{20}, \bibinfo{number}{11} (\bibinfo{year}{1998}).
\newblock

\bibitem[\protect\citeauthoryear{Jiang, Huang, Duan, and Zhao}{Jiang
  et~al\mbox{.}}{2015}]
        {Jiang2015}
\bibfield{author}{\bibinfo{person}{M. Jiang}, \bibinfo{person}{S. Huang},
  \bibinfo{person}{J. Duan}, {and} \bibinfo{person}{Q. Zhao}.}
  \bibinfo{year}{2015}\natexlab{}.
\newblock \showarticletitle{{SALICON: Saliency in Context}}. In
  \bibinfo{booktitle}{\emph{Proc. CVPR}}.
\newblock

\bibitem[\protect\citeauthoryear{Jokinen, Wang, Sarcar, Oulasvirta, and
  Ren}{Jokinen et~al\mbox{.}}{2020}]
        {Jokinen20}
\bibfield{author}{\bibinfo{person}{J.~P. Jokinen}, \bibinfo{person}{Z. Wang},
  \bibinfo{person}{S. Sarcar}, \bibinfo{person}{A. Oulasvirta}, {and}
  \bibinfo{person}{X. Ren}.} \bibinfo{year}{2020}\natexlab{}.
\newblock \showarticletitle{Adaptive feature guidance: Modelling visual search
  with graphical layouts}.
\newblock \bibinfo{journal}{\emph{Int. J. Hum-Comput. Stud.}}
  \bibinfo{volume}{136} (\bibinfo{year}{2020}).
\newblock

\bibitem[\protect\citeauthoryear{Judd, Durand, and Torralba}{Judd
  et~al\mbox{.}}{2012}]
        {Judd12}
\bibfield{author}{\bibinfo{person}{T. Judd}, \bibinfo{person}{F. Durand}, {and}
  \bibinfo{person}{A. Torralba}.} \bibinfo{year}{2012}\natexlab{}.
\newblock \bibinfo{booktitle}{\emph{A Benchmark of Computational Models of
  Saliency to Predict Human Fixations}}.
\newblock \bibinfo{type}{{T}echnical {R}eport}.
\newblock

\bibitem[\protect\citeauthoryear{Judd, Ehinger, Durand, and Torralba}{Judd
  et~al\mbox{.}}{2009}]
        {Ehinger09}
\bibfield{author}{\bibinfo{person}{T. Judd}, \bibinfo{person}{K. Ehinger},
  \bibinfo{person}{F. Durand}, {and} \bibinfo{person}{A. Torralba}.}
  \bibinfo{year}{2009}\natexlab{}.
\newblock \showarticletitle{Learning to predict where humans look}. In
  \bibinfo{booktitle}{\emph{Proc. ICCV}}.
\newblock

\bibitem[\protect\citeauthoryear{Kelly, Miellet, and Caldara}{Kelly
  et~al\mbox{.}}{2010}]
        {Kelly10}
\bibfield{author}{\bibinfo{person}{D.~J. Kelly}, \bibinfo{person}{S. Miellet},
  {and} \bibinfo{person}{R. Caldara}.} \bibinfo{year}{2010}\natexlab{}.
\newblock \showarticletitle{Culture shapes eye movements for visually
  homogeneous objects}.
\newblock \bibinfo{journal}{\emph{Front. Psychol.}}  \bibinfo{volume}{1}
  (\bibinfo{year}{2010}).
\newblock

\bibitem[\protect\citeauthoryear{Kim}{Kim}{2017}]
        {Kim17}
\bibfield{author}{\bibinfo{person}{H.-Y. Kim}.}
  \bibinfo{year}{2017}\natexlab{}.
\newblock \showarticletitle{Statistical notes for clinical researchers:
  {Chi}-squared test and {Fisher}'s exact test}.
\newblock \bibinfo{journal}{\emph{Restor. Dent. Endod.}} \bibinfo{volume}{42},
  \bibinfo{number}{2} (\bibinfo{year}{2017}).
\newblock

\bibitem[\protect\citeauthoryear{Li and Yu}{Li and Yu}{2015}]
        {Li15}
\bibfield{author}{\bibinfo{person}{G. Li} {and} \bibinfo{person}{Y. Yu}.}
  \bibinfo{year}{2015}\natexlab{}.
\newblock \showarticletitle{Visual Saliency Based on Multiscale Deep Features}.
  In \bibinfo{booktitle}{\emph{Proc. CVPR}}.
\newblock

\bibitem[\protect\citeauthoryear{Lindgaard, Fernandes, Dudek, and
  Brown}{Lindgaard et~al\mbox{.}}{2006}]
        {Lindgaard19_web}
\bibfield{author}{\bibinfo{person}{G. Lindgaard}, \bibinfo{person}{G.
  Fernandes}, \bibinfo{person}{C. Dudek}, {and} \bibinfo{person}{J. Brown}.}
  \bibinfo{year}{2006}\natexlab{}.
\newblock \showarticletitle{Attention web designers: You have 50 milliseconds
  to make a good first impression!}
\newblock \bibinfo{journal}{\emph{Behav. Inform. Technol.}}
  \bibinfo{volume}{25}, \bibinfo{number}{2} (\bibinfo{year}{2006}).
\newblock

\bibitem[\protect\citeauthoryear{Liu, Craft, Situ, Yumer, Mech, and Kumar}{Liu
  et~al\mbox{.}}{2018}]
        {Liu:2018:RICO}
\bibfield{author}{\bibinfo{person}{T.~F. Liu}, \bibinfo{person}{M. Craft},
  \bibinfo{person}{J. Situ}, \bibinfo{person}{E. Yumer}, \bibinfo{person}{R.
  Mech}, {and} \bibinfo{person}{R. Kumar}.} \bibinfo{year}{2018}\natexlab{}.
\newblock \showarticletitle{Learning Design Semantics for Mobile Apps}. In
  \bibinfo{booktitle}{\emph{Proc. UIST}}.
\newblock

\bibitem[\protect\citeauthoryear{Long, Cheung, Duong, Paynter, and Asper}{Long
  et~al\mbox{.}}{2017}]
        {Long17}
\bibfield{author}{\bibinfo{person}{J. Long}, \bibinfo{person}{R. Cheung},
  \bibinfo{person}{S. Duong}, \bibinfo{person}{R. Paynter}, {and}
  \bibinfo{person}{L. Asper}.} \bibinfo{year}{2017}\natexlab{}.
\newblock \showarticletitle{Viewing distance and eyestrain symptoms with
  prolonged viewing of smartphones}.
\newblock \bibinfo{journal}{\emph{Clin. Exp. Optom.}} \bibinfo{volume}{100},
  \bibinfo{number}{2} (\bibinfo{year}{2017}).
\newblock

\bibitem[\protect\citeauthoryear{Lu and Lim}{Lu and Lim}{2012}]
        {Lim2012}
\bibfield{author}{\bibinfo{person}{S. Lu} {and} \bibinfo{person}{J.-H. Lim}.}
  \bibinfo{year}{2012}\natexlab{}.
\newblock \showarticletitle{Saliency Modeling from Image Histograms}. In
  \bibinfo{booktitle}{\emph{Proc. ECCV}}.
\newblock

\bibitem[\protect\citeauthoryear{Ma, Hua, Lu, and HJ}{Ma et~al\mbox{.}}{2005}]
        {Ma05}
\bibfield{author}{\bibinfo{person}{Y. Ma}, \bibinfo{person}{X. Hua},
  \bibinfo{person}{L. Lu}, {and} \bibinfo{person}{H.~Z. HJ}.}
  \bibinfo{year}{2005}\natexlab{}.
\newblock \showarticletitle{A generic framework of user attention model and its
  application in video summarization}.
\newblock \bibinfo{journal}{\emph{IEEE Trans. Multimed.}} \bibinfo{volume}{7},
  \bibinfo{number}{5} (\bibinfo{year}{2005}).
\newblock

\bibitem[\protect\citeauthoryear{Marat, Rahman, Pellerin, Guyader, and
  Houzet}{Marat et~al\mbox{.}}{2013}]
        {Marat}
\bibfield{author}{\bibinfo{person}{S. Marat}, \bibinfo{person}{A. Rahman},
  \bibinfo{person}{D. Pellerin}, \bibinfo{person}{N. Guyader}, {and}
  \bibinfo{person}{D. Houzet}.} \bibinfo{year}{2013}\natexlab{}.
\newblock \showarticletitle{Improving Visual Saliency by Adding `Face Feature
  Map' and `Center Bias'}.
\newblock \bibinfo{journal}{\emph{Cogn. Comput.}} \bibinfo{volume}{5},
  \bibinfo{number}{1} (\bibinfo{year}{2013}).
\newblock

\bibitem[\protect\citeauthoryear{Miniukovich and De~Angeli}{Miniukovich and
  De~Angeli}{2014}]
        {Miniukovich14_mobile}
\bibfield{author}{\bibinfo{person}{A. Miniukovich} {and} \bibinfo{person}{A.
  De~Angeli}.} \bibinfo{year}{2014}\natexlab{}.
\newblock \showarticletitle{Visual Impressions of Mobile App Interfaces}. In
  \bibinfo{booktitle}{\emph{Proc. NordiCHI}}.
\newblock

\bibitem[\protect\citeauthoryear{Mishra, Aloimonos, and Fah}{Mishra
  et~al\mbox{.}}{2009}]
        {Mishra09}
\bibfield{author}{\bibinfo{person}{A. Mishra}, \bibinfo{person}{Y. Aloimonos},
  {and} \bibinfo{person}{C. Fah}.} \bibinfo{year}{2009}\natexlab{}.
\newblock \showarticletitle{Active segmentation with fixation}. In
  \bibinfo{booktitle}{\emph{Proc. ICCV}}.
\newblock

\bibitem[\protect\citeauthoryear{Mobile~UI}{Mobile~UI}{2019}]
        {MobileUI:guidelines}
Mobile~UI \bibinfo{year}{2019}\natexlab{}.
\newblock \bibinfo{title}{Mobile {UI} guidelines}.
\newblock \bibinfo{howpublished}{Available: \url{https://mobileui.github.io/}
  (last accessed January 2020)}.
\newblock

\bibitem[\protect\citeauthoryear{{Ni}, {Xu}, {Nguyen}, {Wang}, {Lang}, {Huang},
  and {Yan}}{{Ni} et~al\mbox{.}}{2014}]
        {Ni14touch}
\bibfield{author}{\bibinfo{person}{B. {Ni}}, \bibinfo{person}{M. {Xu}},
  \bibinfo{person}{T.~V. {Nguyen}}, \bibinfo{person}{M. {Wang}},
  \bibinfo{person}{C. {Lang}}, \bibinfo{person}{Z. {Huang}}, {and}
  \bibinfo{person}{S. {Yan}}.} \bibinfo{year}{2014}\natexlab{}.
\newblock \showarticletitle{Touch Saliency: Characteristics and Prediction}.
\newblock \bibinfo{journal}{\emph{IEEE Trans. Multimed.}} \bibinfo{volume}{16},
  \bibinfo{number}{6} (\bibinfo{year}{2014}).
\newblock

\bibitem[\protect\citeauthoryear{Nuthmann and Henderson}{Nuthmann and
  Henderson}{2014}]
        {Nuthmann}
\bibfield{author}{\bibinfo{person}{A. Nuthmann} {and} \bibinfo{person}{J.~M.
  Henderson}.} \bibinfo{year}{2014}\natexlab{}.
\newblock \showarticletitle{Object-based attentional selection in scene
  viewing}.
\newblock \bibinfo{journal}{\emph{J. Vis.}} \bibinfo{volume}{10},
  \bibinfo{number}{8} (\bibinfo{year}{2014}).
\newblock

\bibitem[\protect\citeauthoryear{Ossandon, Onat, and König}{Ossandon
  et~al\mbox{.}}{2014}]
        {Ossandon}
\bibfield{author}{\bibinfo{person}{J.~P. Ossandon}, \bibinfo{person}{S. Onat},
  {and} \bibinfo{person}{P. König}.} \bibinfo{year}{2014}\natexlab{}.
\newblock \showarticletitle{Spatial biases in viewing behavior}.
\newblock \bibinfo{journal}{\emph{J. Vis.}} \bibinfo{volume}{14},
  \bibinfo{number}{2} (\bibinfo{year}{2014}).
\newblock

\bibitem[\protect\citeauthoryear{Ouerhani, Bracamonte, Hugli, Ansorge, and
  Pellandini}{Ouerhani et~al\mbox{.}}{2001}]
        {Ouerhani01}
\bibfield{author}{\bibinfo{person}{N. Ouerhani}, \bibinfo{person}{J.
  Bracamonte}, \bibinfo{person}{H. Hugli}, \bibinfo{person}{M. Ansorge}, {and}
  \bibinfo{person}{F. Pellandini}.} \bibinfo{year}{2001}\natexlab{}.
\newblock \showarticletitle{Adaptive color image compression based on visual
  attention}. In \bibinfo{booktitle}{\emph{Proc. ICIAP}}.
\newblock

\bibitem[\protect\citeauthoryear{Parkhurst, Law, and Niebur}{Parkhurst
  et~al\mbox{.}}{2002}]
        {Parkhurst02}
\bibfield{author}{\bibinfo{person}{D. Parkhurst}, \bibinfo{person}{K. Law},
  {and} \bibinfo{person}{E. Niebur}.} \bibinfo{year}{2002}\natexlab{}.
\newblock \showarticletitle{Modeling the role of salience in the allocation of
  overt visual attention}.
\newblock \bibinfo{journal}{\emph{Vis. Res.}} \bibinfo{volume}{42},
  \bibinfo{number}{1} (\bibinfo{year}{2002}).
\newblock

\bibitem[\protect\citeauthoryear{Rayner, Liversedge, Nuthmann, Kliegl, and
  G.}{Rayner et~al\mbox{.}}{2009}]
        {Rayner}
\bibfield{author}{\bibinfo{person}{K. Rayner}, \bibinfo{person}{S.~P.
  Liversedge}, \bibinfo{person}{A. Nuthmann}, \bibinfo{person}{R. Kliegl},
  {and} \bibinfo{person}{U. G.}} \bibinfo{year}{2009}\natexlab{}.
\newblock \showarticletitle{Rayner’s 1979 paper}.
\newblock \bibinfo{journal}{\emph{Perception}} \bibinfo{volume}{38},
  \bibinfo{number}{6} (\bibinfo{year}{2009}).
\newblock

\bibitem[\protect\citeauthoryear{Rosenholtz, Dorai, and Freeman}{Rosenholtz
  et~al\mbox{.}}{2011}]
        {Rosenholtz11}
\bibfield{author}{\bibinfo{person}{R. Rosenholtz}, \bibinfo{person}{A. Dorai},
  {and} \bibinfo{person}{R. Freeman}.} \bibinfo{year}{2011}\natexlab{}.
\newblock \showarticletitle{Do Predictions of Visual Perception Aid Design?}
\newblock \bibinfo{journal}{\emph{ACM Trans. Appl. Percept.}}
  \bibinfo{volume}{8}, \bibinfo{number}{2} (\bibinfo{year}{2011}).
\newblock

\bibitem[\protect\citeauthoryear{Semantic~UI}{Semantic~UI}{2019}]
        {SemanticUI:guidelines}
Semantic~UI \bibinfo{year}{2019}\natexlab{}.
\newblock \bibinfo{title}{Semantic {UI} guidelines}.
\newblock \bibinfo{howpublished}{Available: \url{https://semantic-ui.com} (last
  accessed January 2020)}.
\newblock

\bibitem[\protect\citeauthoryear{Seri{\`e}s and Seitz}{Seri{\`e}s and
  Seitz}{2013}]
        {series2013learning}
\bibfield{author}{\bibinfo{person}{P. Seri{\`e}s} {and} \bibinfo{person}{A.
  Seitz}.} \bibinfo{year}{2013}\natexlab{}.
\newblock \showarticletitle{Learning what to expect (in visual perception)}.
\newblock \bibinfo{journal}{\emph{Front. Hum. neurosci.}}  \bibinfo{volume}{7}
  (\bibinfo{year}{2013}).
\newblock

\bibitem[\protect\citeauthoryear{Shelhamer, Long, and Darrell}{Shelhamer
  et~al\mbox{.}}{2017}]
        {Shelhamer17}
\bibfield{author}{\bibinfo{person}{E. Shelhamer}, \bibinfo{person}{J. Long},
  {and} \bibinfo{person}{T. Darrell}.} \bibinfo{year}{2017}\natexlab{}.
\newblock \showarticletitle{Fully Convolutional Networks for Semantic
  Segmentation}.
\newblock \bibinfo{journal}{\emph{IEEE Trans. Pattern Anal. Mach. Intell.}}
  \bibinfo{volume}{39}, \bibinfo{number}{4} (\bibinfo{year}{2017}).
\newblock

\bibitem[\protect\citeauthoryear{Shen and Zhao}{Shen and Zhao}{2014}]
        {Shen14}
\bibfield{author}{\bibinfo{person}{C. Shen} {and} \bibinfo{person}{Q. Zhao}.}
  \bibinfo{year}{2014}\natexlab{}.
\newblock \showarticletitle{Webpage Saliency}. In
  \bibinfo{booktitle}{\emph{Proc. ECCV}}.
\newblock

\bibitem[\protect\citeauthoryear{Siagian and Itti}{Siagian and Itti}{2007}]
        {Siagian07}
\bibfield{author}{\bibinfo{person}{C. Siagian} {and} \bibinfo{person}{L.
  Itti}.} \bibinfo{year}{2007}\natexlab{}.
\newblock \showarticletitle{Rapid biologically-inspired scene classification
  using features shared with visual attention}.
\newblock \bibinfo{journal}{\emph{IEEE Trans. Pattern Anal. Mach. Intell.}}
  \bibinfo{volume}{29}, \bibinfo{number}{2} (\bibinfo{year}{2007}).
\newblock

\bibitem[\protect\citeauthoryear{Still and Masciocchi}{Still and
  Masciocchi}{2010}]
        {Still10}
\bibfield{author}{\bibinfo{person}{J.~D. Still} {and} \bibinfo{person}{C.~M.
  Masciocchi}.} \bibinfo{year}{2010}\natexlab{}.
\newblock \showarticletitle{A Saliency Model Predicts Fixations in Web
  Interfaces}. In \bibinfo{booktitle}{\emph{Proc. MDDAUI Workshop}}.
\newblock

\bibitem[\protect\citeauthoryear{Tatler}{Tatler}{2007}]
        {Tatler2007}
\bibfield{author}{\bibinfo{person}{B.~W. Tatler}.}
  \bibinfo{year}{2007}\natexlab{}.
\newblock \showarticletitle{{The central fixation bias in scene viewing:
  selecting an optimal viewing position independently of motor bases and image
  feature distributions.}}
\newblock \bibinfo{journal}{\emph{J. Vis.}} \bibinfo{volume}{14},
  \bibinfo{number}{7} (\bibinfo{year}{2007}).
\newblock

\bibitem[\protect\citeauthoryear{Tavakoli, Ahmed, Borji, and
  Laaksonen}{Tavakoli et~al\mbox{.}}{2017}]
        {Tavakoli_2017_CVPR}
\bibfield{author}{\bibinfo{person}{H.~R. Tavakoli}, \bibinfo{person}{F. Ahmed},
  \bibinfo{person}{A. Borji}, {and} \bibinfo{person}{J. Laaksonen}.}
  \bibinfo{year}{2017}\natexlab{}.
\newblock \showarticletitle{Saliency Revisited: Analysis of Mouse Movements
  Versus Fixations}. In \bibinfo{booktitle}{\emph{Proc. CVPR}}.
\newblock

\bibitem[\protect\citeauthoryear{Tieleman and Hinton}{Tieleman and
  Hinton}{2012}]
        {tieleman2012lecture}
\bibfield{author}{\bibinfo{person}{T. Tieleman} {and} \bibinfo{person}{G.
  Hinton}.} \bibinfo{year}{2012}\natexlab{}.
\newblock \showarticletitle{{RMSProp}: Divide the gradient by a running average
  of its recent magnitude}.
\newblock \bibinfo{journal}{\emph{Coursera: Neural networks for machine
  learning}} \bibinfo{volume}{4}, \bibinfo{number}{2} (\bibinfo{year}{2012}).
\newblock

\bibitem[\protect\citeauthoryear{Tseng, Carmi, Cameron, Munoz, and Itti}{Tseng
  et~al\mbox{.}}{2009}]
        {Tseng09}
\bibfield{author}{\bibinfo{person}{P.-H. Tseng}, \bibinfo{person}{R. Carmi},
  \bibinfo{person}{I.~G. Cameron}, \bibinfo{person}{D.~P. Munoz}, {and}
  \bibinfo{person}{L. Itti}.} \bibinfo{year}{2009}\natexlab{}.
\newblock \showarticletitle{Quantifying center bias of observers in free
  viewing of dynamic natural scenes}.
\newblock \bibinfo{journal}{\emph{J. Vis.}} \bibinfo{volume}{9},
  \bibinfo{number}{7} (\bibinfo{year}{2009}).
\newblock

\bibitem[\protect\citeauthoryear{Tsotsos}{Tsotsos}{1991}]
        {Tsotsos91}
\bibfield{author}{\bibinfo{person}{J.~K. Tsotsos}.}
  \bibinfo{year}{1991}\natexlab{}.
\newblock \showarticletitle{Is Complexity Theory appropriate for analysing
  biological systems?}
\newblock \bibinfo{journal}{\emph{Behav. Brain Sci.}} \bibinfo{volume}{14},
  \bibinfo{number}{4} (\bibinfo{year}{1991}).
\newblock

\bibitem[\protect\citeauthoryear{Veale, Hafed, and Yoshida}{Veale
  et~al\mbox{.}}{2017a}]
        {Richard}
\bibfield{author}{\bibinfo{person}{R. Veale}, \bibinfo{person}{Z. Hafed}, {and}
  \bibinfo{person}{M. Yoshida}.} \bibinfo{year}{2017}\natexlab{a}.
\newblock \showarticletitle{How is visual salience computed in the brain?
  Insights from behaviour, neurobiology and modelling}.
\newblock \bibinfo{journal}{\emph{Phil. Trans. R. Soc. B.}} \bibinfo{number}{1}
  (\bibinfo{year}{2017}).
\newblock

\bibitem[\protect\citeauthoryear{Veale, Hafed, and Yoshida}{Veale
  et~al\mbox{.}}{2017b}]
        {Veale17}
\bibfield{author}{\bibinfo{person}{R. Veale}, \bibinfo{person}{Z.~M. Hafed},
  {and} \bibinfo{person}{M. Yoshida}.} \bibinfo{year}{2017}\natexlab{b}.
\newblock \showarticletitle{How is visual salience computed in the brain?
  Insights from behaviour, neurobiology and modelling}.
\newblock \bibinfo{journal}{\emph{Philos. Trans. R. Soc. Lond. B. Biol. Sci.}}
  \bibinfo{volume}{372}, \bibinfo{number}{1714} (\bibinfo{year}{2017}).
\newblock

\bibitem[\protect\citeauthoryear{Vidyapu, Vedula, and Bhattacharya}{Vidyapu
  et~al\mbox{.}}{2019}]
        {Vidyapu19}
\bibfield{author}{\bibinfo{person}{S. Vidyapu}, \bibinfo{person}{V.~S. Vedula},
  {and} \bibinfo{person}{S. Bhattacharya}.} \bibinfo{year}{2019}\natexlab{}.
\newblock \showarticletitle{Quantitative Visual Attention Prediction on Webpage
  Images Using Multiclass SVM}. In \bibinfo{booktitle}{\emph{Proc. ETRA}}.
\newblock

\bibitem[\protect\citeauthoryear{Wang and Pomplun}{Wang and Pomplun}{2012}]
        {Wang2012}
\bibfield{author}{\bibinfo{person}{H.-C. Wang} {and} \bibinfo{person}{M.
  Pomplun}.} \bibinfo{year}{2012}\natexlab{}.
\newblock \showarticletitle{The attraction of visual attention to texts in
  real-world scenes}.
\newblock \bibinfo{journal}{\emph{J. Vis.}} \bibinfo{volume}{12},
  \bibinfo{number}{26} (\bibinfo{year}{2012}).
\newblock

\bibitem[\protect\citeauthoryear{Wolfe and Horowitz}{Wolfe and
  Horowitz}{2004}]
        {Wolfe04}
\bibfield{author}{\bibinfo{person}{J.~M. Wolfe} {and} \bibinfo{person}{T.~S.
  Horowitz}.} \bibinfo{year}{2004}\natexlab{}.
\newblock \showarticletitle{What attributes guide the deployment of visual
  attention and how do they do it?}
\newblock \bibinfo{journal}{\emph{Nat. Rev. Neurosci.}} \bibinfo{volume}{5},
  \bibinfo{number}{6} (\bibinfo{year}{2004}).
\newblock

\bibitem[\protect\citeauthoryear{Xu, Ni, Dong, Huang, Wang, and Yan}{Xu
  et~al\mbox{.}}{2012}]
        {Xu12touch}
\bibfield{author}{\bibinfo{person}{M. Xu}, \bibinfo{person}{B. Ni},
  \bibinfo{person}{J. Dong}, \bibinfo{person}{Z. Huang}, \bibinfo{person}{M.
  Wang}, {and} \bibinfo{person}{S. Yan}.} \bibinfo{year}{2012}\natexlab{}.
\newblock \showarticletitle{Touch Saliency}. In \bibinfo{booktitle}{\emph{Proc.
  ACM Multimedia}}.
\newblock

\bibitem[\protect\citeauthoryear{Xu, Sugano, and Bulling}{Xu
  et~al\mbox{.}}{2016}]
        {Xu16_chi}
\bibfield{author}{\bibinfo{person}{P. Xu}, \bibinfo{person}{Y. Sugano}, {and}
  \bibinfo{person}{A. Bulling}.} \bibinfo{year}{2016}\natexlab{}.
\newblock \showarticletitle{Spatio-Temporal Modeling and Prediction of Visual
  Attention in Graphical User Interfaces}. In \bibinfo{booktitle}{\emph{Proc.
  CHI}}.
\newblock

\bibitem[\protect\citeauthoryear{Zhang and Sclaroff}{Zhang and
  Sclaroff}{2013}]
        {Zhang13}
\bibfield{author}{\bibinfo{person}{J. Zhang} {and} \bibinfo{person}{S.
  Sclaroff}.} \bibinfo{year}{2013}\natexlab{}.
\newblock \showarticletitle{Saliency Detection: A Boolean Map Approach}. In
  \bibinfo{booktitle}{\emph{Proc. ICCV}}.
\newblock

\bibitem[\protect\citeauthoryear{Zhao and Koch}{Zhao and Koch}{2013}]
        {Zhao13}
\bibfield{author}{\bibinfo{person}{Q. Zhao} {and} \bibinfo{person}{C. Koch}.}
  \bibinfo{year}{2013}\natexlab{}.
\newblock \showarticletitle{Learning saliency-based visual attention: A
  review}.
\newblock \bibinfo{journal}{\emph{Signal Process.}}  \bibinfo{volume}{93}
  (\bibinfo{year}{2013}).
\newblock

\end{thebibliography}
\end{document}